\RequirePackage{fix-cm}
\documentclass[natbib,smallextended]{svjour3}       
\smartqed  
\usepackage{graphicx,amsmath,amssymb}
 \newcommand{\be}{\begin{equation}}
\newcommand{\ee}{\end{equation}}
\newcommand{\beq}{\begin{eqnarray}}
\newcommand{\eeq}{\end{eqnarray}}
\newcommand\subsun[1]{{$_{\normalsize\odot}$}}

\def\lsim{\;\raise0.3ex\hbox{$<$\kern-0.75em\raise-1.1ex\hbox{$\approx  $}}\;}
\def\gsim{\;\raise0.3ex\hbox{$>$\kern-0.75em\raise-1.1ex\hbox{$\sim $}}\;}

\newcommand{\ion}[2]{#1\,{\sc{#2}}}

\begin{document}

\title{Hot Atmospheres, Cold Gas, AGN Feedback and the Evolution of Early Type Galaxies: a Topical Perspective}



\titlerunning{Hot Atmospheres and AGN Feedback in ETGs}        

\author{N. Werner         \and
        B. R. McNamara	\and
        E. Churazov		\and
        E. Scannapieco 
}

\authorrunning{Werner et al.} 

\institute{N. Werner \at
            MTA-E\"otv\"os University Lend\"ulet Hot Universe and Astrophysics  Research Group, P\'azm\'any P\'eter s\'et\'any 1/A, Budapest, 1117, Hungary \\
            Department of Theoretical Physics and Astrophysics, Faculty of Science, Masaryk University, Kotl\'a\v{r}sk\'a 2, Brno, 611 37, Czech Republic \\
            School of Science, Hiroshima University, 1-3-1 Kagamiyama, Higashi-Hiroshima 739-8526, Japan \\
    \email{wernernorbi@gmail.com}           
           \and
           B. R. McNamara \at
              Department of Physics and Astronomy, University of Waterloo, 200 University Ave W, Waterloo, ON N2L 3G1, Canada
           \and
           E. Churazov \at
           Max-Planck-Institut fur Astrophysik, Karl-Schwarzschild-Strasse 1, Garching 85741, Germany \\
        Space Research Institute, Profsoyuznaya 84/32, Moscow 117997, Russia
           \and
           E. Scannapieco \at
           School of Earth and Space Exploration, Arizona State University, P.O. Box 871404, Tempe, AZ 85287-1404.}

\date{Received: date / Accepted: date}

\maketitle

\begin{abstract}

Most galaxies comparable to or larger than the mass of the Milky Way host hot, X-ray emitting atmospheres, and many such galaxies are radio sources. Hot atmospheres and radio jets and lobes are the ingredients of radio-mechanical active galactic nucleus (AGN) feedback. While a consensus has emerged that such feedback suppresses cooling of hot cluster atmospheres, less attention has been paid to massive galaxies where similar mechanisms are at play. Observation indicates that the atmospheres of elliptical and S0 galaxies were accreted externally during the process of galaxy assembly and augmented significantly by stellar mass loss. 
Their atmospheres have entropy and cooling time profiles that are remarkably similar to those of central cluster galaxies.  About half display filamentary or disky nebulae of cool and cold gas, much of which has likely cooled from the hot atmospheres. We review the observational and theoretical perspectives on thermal instabilities in galactic atmospheres and the evidence that AGN heating is able to roughly balance the atmospheric cooling. Such heating and cooling may be regulating star formation in all massive spheroids at late times.


\end{abstract}

\section{Introduction}
\label{intro}

Elliptical galaxies are transitional systems in cosmological structure formation.   Lower mass galaxies are composed of star forming disks embedded in a galaxy with a single dark matter halo. At larger masses, galaxies are spheroids with little star formation, many of which populate a single dark matter halo.  At $z=0$ many of the largest systems are in galaxy clusters, in which most of the baryonic mass is located not within galaxies themselves, but rather within a diffuse, hot medium. 

Historically, this transition point was understood in the context of cooling in the dark matter structure formation model.  The collapse of baryonic matter follows the collapse of overdense regions of dark matter \citep{white1978,whitefrenk1991,kauffmann1993,lacey1993}. These overdense regions grow over time by accretion and merging.  As baryons fall into growing dark matter halos, strong shocks form and heat them. This heat must be radiated away before star formation can occur. Radiative cooling is less efficient in more massive halos, which form later and have higher virial temperatures. Therefore, in this picture, elliptical galaxies and galaxy clusters are mostly quiescent because the bulk of their atmospheres have not had time to cool \cite[e.g.][]{binney1977,rees1977,silk1977}.

Several lines of inquiry have since uncovered problems with this picture.  As cosmological simulations improved, they revealed that cooling is significantly enhanced by gas inhomogeneities, allowing galaxies to form in the simulations with stellar masses over 10 times larger than observed \cite[][]{Suginohara1998,dave2001}. At the same time, observations of the diffuse medium in galaxy clusters and groups showed that their X-ray luminosities and temperatures are not related as $L_{\rm X} \propto T^2$, as they would be if heating were purely gravitational \citep{kaiser1991,evrard1991}.  Instead, the slope of this relation steepens considerably at lower temperatures \cite[e.g.][]{david1993,arnaud1999,helsdon2000}.

Galaxy surveys soon revealed that the formation history of large galaxies was also in conflict with theoretical expectations.  Most models predicted that at late cosmological times, when the gas would have had time to cool, massive galaxies would form stars efficiently. Instead, the largest galaxies were apparently dormant by $z \approx   2$ and the smaller ones continued to form stars at much lower redshifts \citep{fontana2004,glazebrook2004,vandokkum2004,arnouts2005,treu2005}.

It was realised that this ``antihierarchical'' trend could be explained naturally in models that include significant heating associated with outflows (jets and winds driven by active galactic nuclei and supernovae) from large galaxies \citep[][]{scannapieco2004,binney2004,granato2004,scannapieco2005,croton2006,thacker2006,dimatteo2008}. In this picture, energetic outflows heat the surrounding medium to temperatures high enough to prevent it from cooling and forming stars. This feedback requires an energetic outflow driven by a large galaxy to be effective in the dense, high-redshift universe.  In the more tenuous, low-redshift universe, equivalently long cooling times can be achieved by less energetic winds. This causes feedback to become more efficient in smaller galaxies, at lower redshifts.

Assuming the large energy demands could be met, heating would not only quench star formation in large galaxies, but could also explain trends observed in clusters of galaxies.  While starburst galaxies are observed to host massive supernova-driven outflows \citep[e.g.][]{pettini2001,veilleux2005,weiner2009,strickland2009,martin2013,chisholm2017},  they are not able to generate enough energy to account for the observed trends  \citep[e.g.][]{cavaliere1998,balogh1999,brighenti2001,babul2002}.  

\begin{figure}
\begin{center}
\includegraphics[width=12cm]{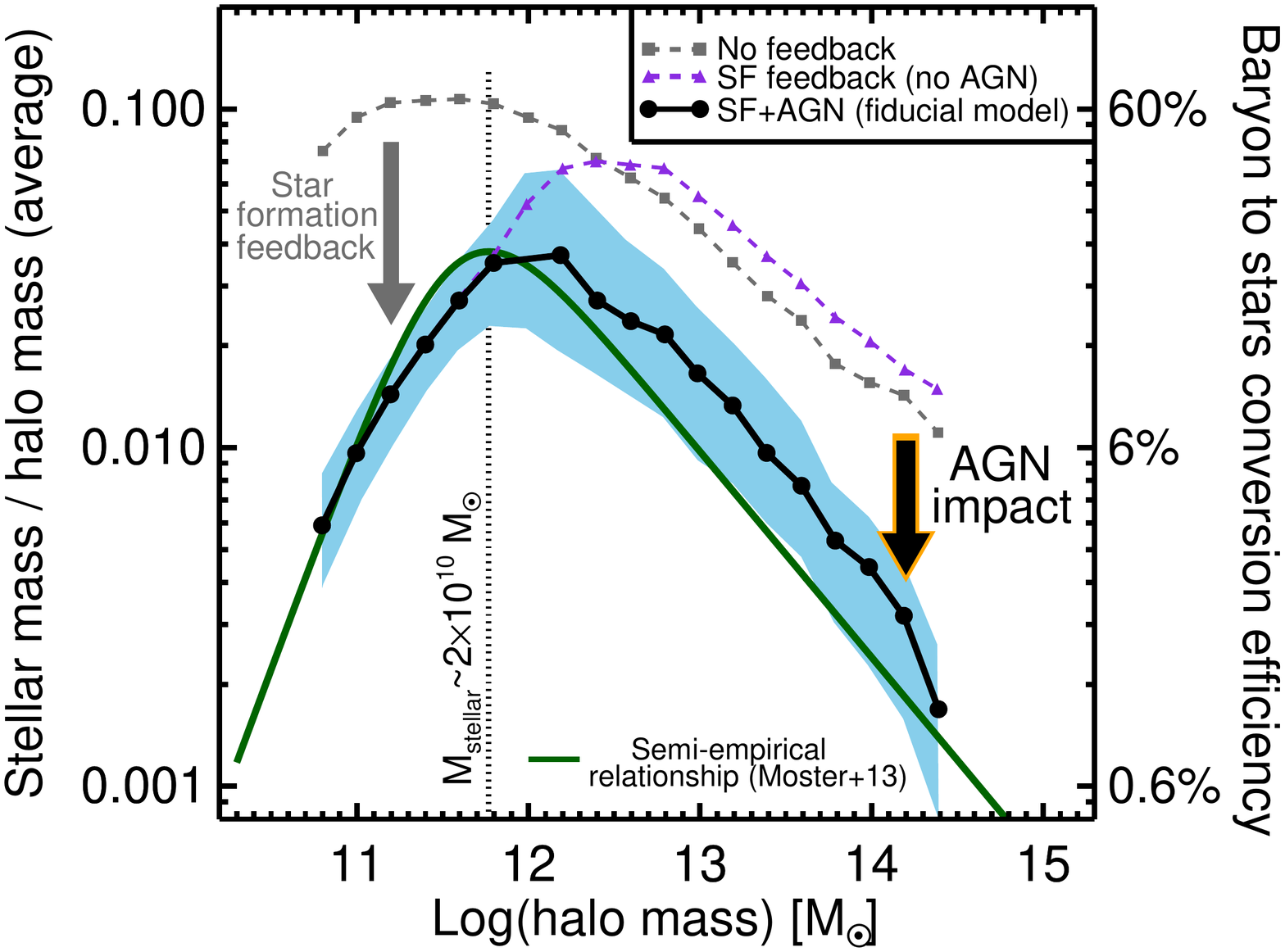}
\end{center}
\caption{The ratio of stellar mass to halo mass as a function of halo mass from the review by \citet{harrison2017} for three different simulation runs by \citet{somerville2008} and for the semi-empirical relation by \citet{moster2013}. The shaded region shows the 16th and 84th percentiles for the fiducial model of \citet{somerville2008} that includes energy injection from AGN and star formation (SF). The y-axis on the right side shows the efficiency for turning baryons into stars ($M_\star/[f_{\rm b} M_{\rm halo}]$, where $f_{\rm b}  = 0.17$ is the cosmological baryon fraction). While stellar feedback reduces the efficiency of converting baryons into stars in low mass haloes, massive haloes require energy injection from AGN.}
\label{fig1}       
\end{figure}

Accreting black holes can generate enough energy \citep{scannapieco2004,granato2004,croton2006,thacker2006, magorrian1998}, as they are the most efficient engines in the universe at converting rest-mass into energy, releasing $\sim 10^{20}~\rm erg$ per gram of accreted gas. This energy may be released in a radiative or mechanical form depending on the accretion rate and structure of the accretion flow. In order to reproduce the observed properties of galaxies, virtually all galaxy formation models require Active Galactic Nuclei (AGN) to heat and/or expel the gas from massive galaxies. AGN feedback effectively reduces the efficiency of converting baryons into stars in systems with stellar masses above $M_{\rm stellar}\approx  2\times10^{10}~M_{\odot}$ (see Fig.~\ref{fig1}). 

AGN dominated by radiation usually do shine in the X-ray, UV, optical and infrared bands.  They are most common in galaxies with ongoing star formation and relatively young stellar populations. Their energy is released in a ``quasar'' or ``wind'' mode associated with accretion rates exceeding $\approx   1\%$ of the Eddington rate. Heating by radiation-driven winds during the high-accretion, quasar-phase of evolution \citep[e.g.][]{silk1998} may play an essential role. However, estimates of the kinetic energy input during this phase vary broadly: from $1\%$ \citep{dekool2001} through 5\% \citep{borguet2013}, up to 60\% of the total luminous energy \citep{chartas2007}. 

Mechanical energy input, dubbed ``radio'' or ``mechanical'' mode, is associated with low Eddington accretion rates. Radio mode, or radio-mechanical AGN are generally found in massive systems surrounded by hot X-ray atmospheres.  Radio jets and lobes of synchrotron emitting plasma couple efficiently to the volume-filling hot atmospheres and are energetically able to balance cooling in the cores of large galaxy clusters \citep{churazov2000,mcnamara2000}. This mechanism is almost certainly regulating star formation in central cluster galaxies. We suggest here that radio mode AGN are likely keeping star formation at low levels in all massive galaxies.

\section{Formation of Hot Atmospheres Surrounding Massive Galaxies}

\subsection{Giant Elliptical Galaxies and Their Progenitors}

According to current models, the progenitors of giant elliptical galaxies formed early in the most massive dark matter haloes \citep[e.g.][]{cattaneo2009}. Gas streamed towards the centres of halos forming the cores of the galaxies and their central, massive black holes \citep[e.g.][]{khalatyan2008}. They apparently grew quickly by efficiently dissipating their energy and angular momentum. Above a critical mass, $M_{\rm crit} \approx  10^{12}~M_\odot$ \citep[e.g.][]{correa2018}, the cooling time of gas at the halo's virial temperature exceeds its free-fall time, causing accretion onto the central galaxy to slow dramatically \citep[e.g.][]{whitefrenk1991}. This accreting gas, heated by shock waves, formed the first X-ray-emitting atmospheres.  

Hubble deep field observations indicate that primeval massive galaxies were compact, approximately four times smaller than the massive, modern galaxies \citep{daddi2005,trujillo2006,trujillo2007}. These so called ``red nuggets,'' with effective radii $r_{\rm e}\lesssim2$ kpc and stellar masses $M_\star\gtrsim10^{11}~M_\odot$ are typically observed at or beyond redshift two. Their rapid early growth presumably preceded a slow accretion phase of numerous stripping encounters and dry mergers that frosted the galaxies with stars in their outer regions.  Their centres were mostly unaffected. But their overall sizes increased dramatically, transforming them into modern giant ellipticals \citep[e.g.][]{hopkins2009,oser2012}. 

If this picture is correct, red nuggets must harbour hot atmospheres. Unfortunately, the sensitivities of our current X-ray observatories do not allow us to observe such atmospheres around high redshift galaxies. In fact, even future X-ray observatories with large effective areas, such as {\it Athena}, will only be able to observe the X-ray atmospheres of massive galaxies out to redshifts $\lesssim 1$. 

However, due to the stochastic nature of mergers, a few red nuggets must have avoided the second stage of growth, remaining almost unaltered since their formation \citep{quilis2013}. The first confirmed low redshift massive relic galaxy, mimicking the properties of high-redshift red nuggets is NGC~1277 in the Perseus cluster \citep{trujillo2014}, which stripped most of its hot X-ray emitting atmosphere. Recently, \citet{ferre2017} confirmed two other ``red nuggets'' in the present day Universe:  MRK 1216 ($D = 97$ Mpc) and PGC 032873 ($D = 108$ Mpc). The stellar masses of these galaxies reach $M_{\star}\approx   2\times10^{11}~M_{\odot}$ and their stellar populations are highly centrally concentrated, resulting in compact morphologies ($R_{\rm e}\sim  2$~kpc) with no signs of interactions. Importantly, the closest neighbours of these galaxies lie at distances $\gtrsim1$~Mpc \citep{ferre2017,yildirim2017}. Therefore, these massive relic systems provide the best opportunity to test the formation models and study the hot atmospheres of the progenitors of giant ellipticals.

\begin{figure}
\begin{center}
\includegraphics[width=11cm]{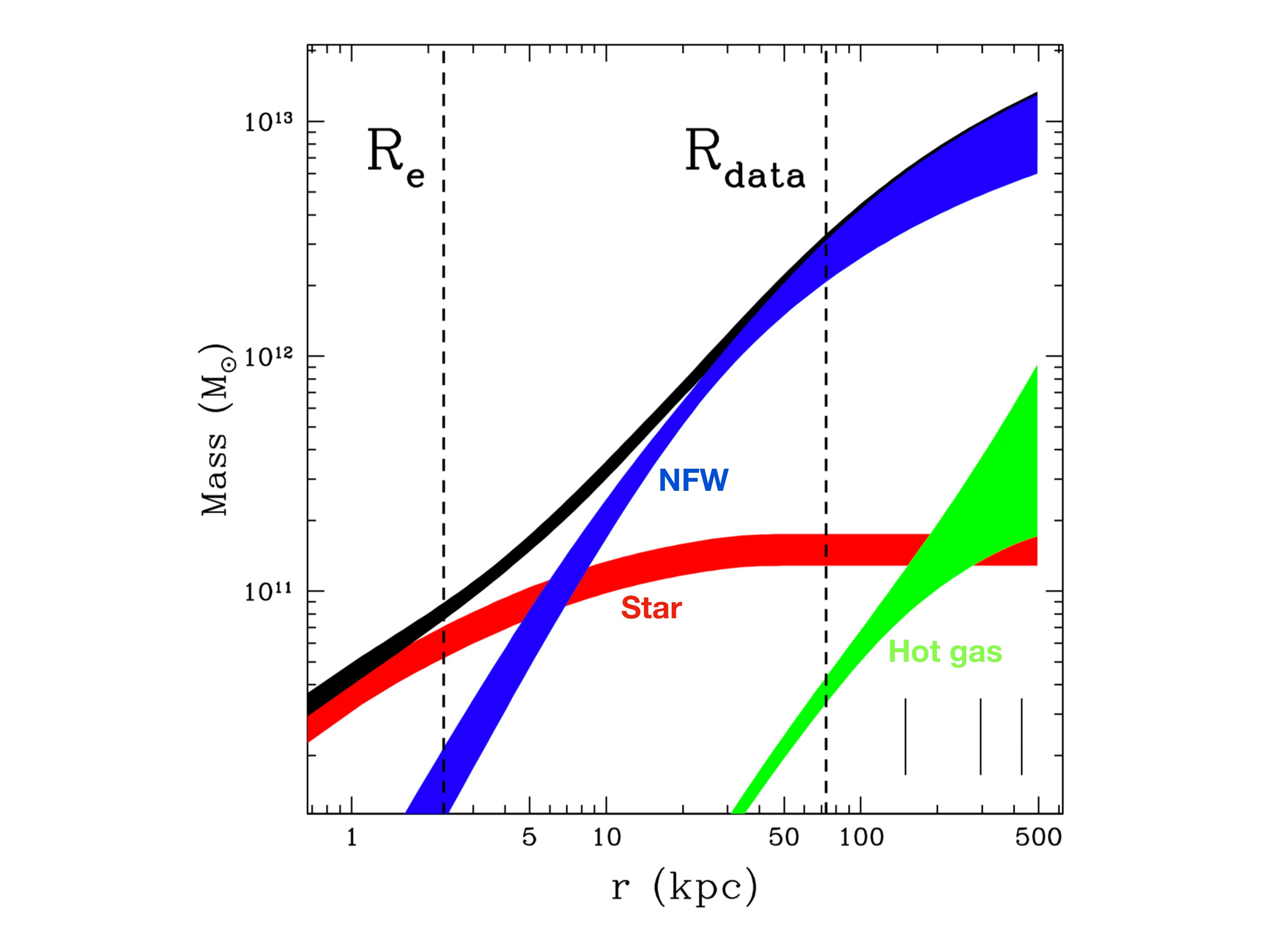}
\end{center}
\caption{Radial distributions of the total mass (black), NFW \citep{navarro1997} dark matter (blue), stellar mass (red), and hot gas mass (green) for the massive, compact, relic galaxy MRK 1216. These relative masses are also typical for giant ellipticals. The black vertical lines in the bottom right corner indicate $r_{2500}$, $r_{500}$, and $r_{200}$. The vertical dashed lines indicate the location of the stellar half-light radius ($R_{\rm e}$) and the outer extent of the X-ray data ($R_{\rm data}$). From \citet{buote2018}.}
\label{buote}       
\end{figure}

\citet{werner2018} and \citet{buote2018} discovered hot X-ray emitting atmospheres around these ``local red nuggets'', which extend far beyond their stellar populations. The atmosphere of MRK 1216 extends out to $r\gtrsim55$ kpc and has an 0.5--7 keV X-ray luminosity of $L_{\rm X}=(7.0\pm0.2)\times10^{41}$~erg~s$^{-1}$, which is similar to the nearby X-ray bright giant ellipticals. The  total mass of MRK~1216 calculated under the assumption of a hydrostatic equilibrium in the hot atmosphere is $M_{200} = (9.6\pm3.7)\times10^{12}~M_\odot$ \citep[see Fig.~\ref{buote},][]{buote2018}, which is about an order of magnitude higher than $M_{\rm crit} \sim 10^{12}~M_\odot$. 

While the mean mass weighted age of MRK~1216 is $12.8\pm1.5$~Gyr, with 99\% of the stellar population more than 10~Gyrs old \citep{ferre2017}, the central cooling time of the X-ray emitting atmosphere is only $t_{\rm cool}=52\pm5$~Myr \citep{werner2018}. The presence of an X-ray atmosphere with a short nominal cooling time and the lack of young stars indicate the presence of a sustained heating source, which prevented star formation since the dissipative formation of the galaxy 13 Gyrs ago. Furthermore, the central temperature peak and the presence of radio emission in the galactic nucleus indicate that the heating source is radio-mode AGN feedback. 

Despite its similar mass \citep{buote2018}, PGC 032873 is an order of magnitude fainter in X-rays with $L_{\rm X} =  (5.6\pm0.5)\times10^{40}$~erg~s$^{-1}$. Given that both MRK 1216 and PGC 032873 appear to have evolved in isolation, the order of magnitude difference in their current X-ray luminosity could be traced back to a difference in the ferocity of their AGN outbursts. 

\begin{figure}
\begin{center}
\includegraphics[width=11cm]{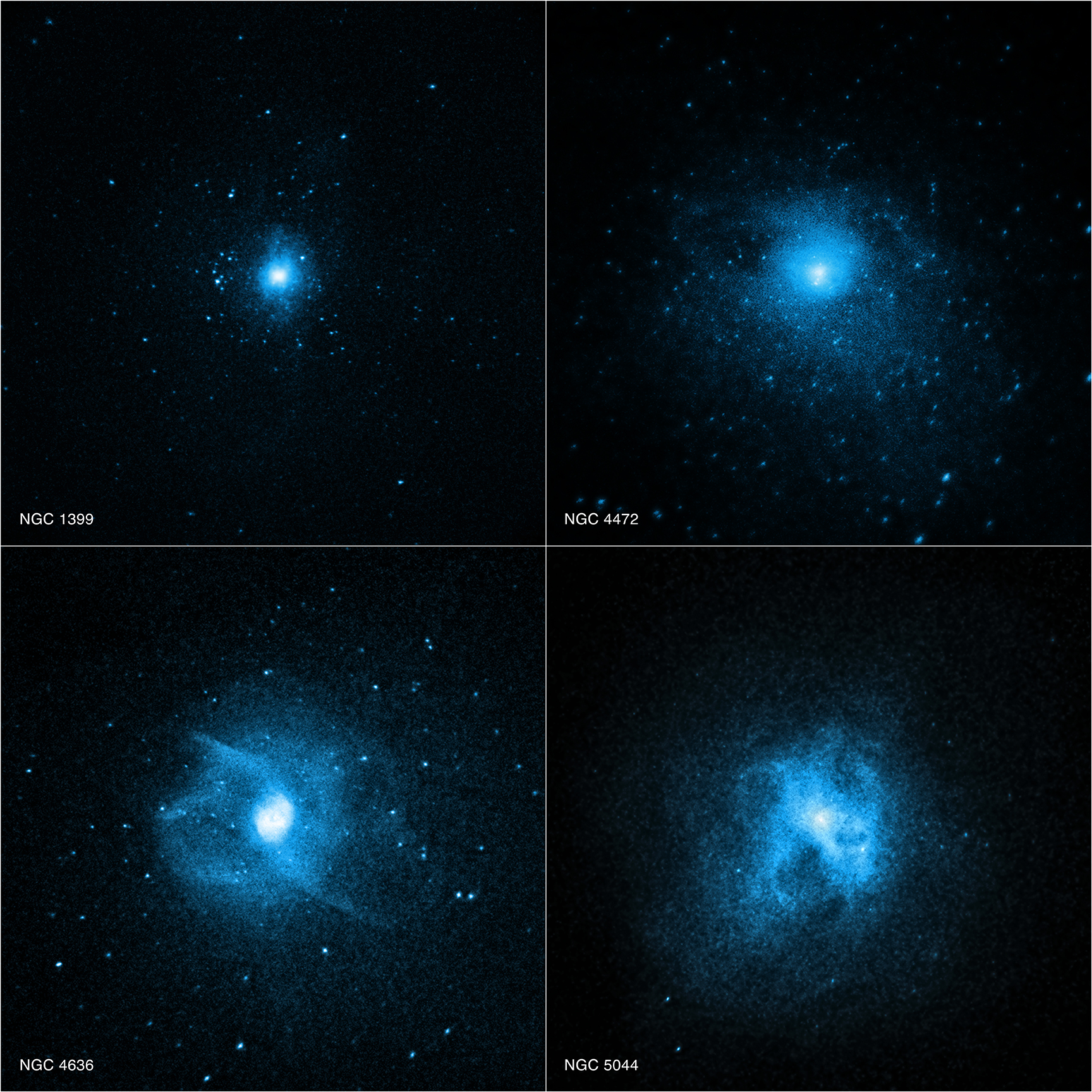}
\end{center}
\caption{{\it Chandra} X-ray images of giant elliptical galaxies reveal hot X-ray bright atmospheres, which are often visibly perturbed by the central AGN (NASA/CXC/Stanford Univ/N. Werner et al. 2014). }
\label{fig2}       
\end{figure}

Giant ellipticals that evolved from early red nuggets today are slowly-rotating, triaxial, and boxy. These characteristics likely emerged from many dry mergers that canceled the angular momenta of the merging galaxies \citep{Kormendy2013}. The most massive have faint cores--- missing light--- with respect to the inward extrapolation of the outer S\'ersic profiles \citep{kormendy2009}. The cores may have been excavated by binary black holes which flung stars out to larger radii. Virtually all of these massive galaxies harbour hot atmospheres. Many have nuclear radio sources \citep[e.g.][]{dunn2010}. X-ray images of radio-bright giant ellipticals often reveal visibly perturbed atmospheres (see Fig.~\ref{fig2}).

Early spirals, S0s, and rapidly-rotating, coreless, flattened ellipticals host lower luminosity X-ray atmospheres \citep{sarzi2013}. The abundance ratios of their stellar populations indicate star formation histories substantially different from those of massive, cored, giant ellipticals. \citet{conroy2014} found relative abundances of $\alpha$ elements, such as O, Mg, and Si, with respect to Fe, to increase as a function of stellar velocity dispersion and stellar mass. The standard interpretation of the [$\alpha$/Fe] ratio is that it is sensitive to the timescale of star formation. Higher values correspond to shorter timescales. The stellar populations of massive early type galaxies with $\sigma\gtrsim200$~km~s$^{-1}$ have typically high [$\alpha$/Fe] ratios, indicating star-formation timescales of $<1$~Gyr and perhaps as short as 0.2 Gyr \citep{conroy2014}.
On the other hand, flattened coreless ellipticals and S0 galaxies, have lower [$\alpha$/Fe] abundance ratios, which indicate that they continued to grow and form stars over billions of years \citep{thomas2005}. 

Fast-rotating coreless ellipticals and S0 galaxies likely evolved via wet mergers, which led to the formation of rotating disks. Among their likely progenitors are the Ultra-Luminous Infra-Red Galaxies (ULIRG): rapidly star-forming, dusty, mergers with infrared luminosities above $10^{12}~L_{\odot}$. The
structural parameters of these galaxies are consistent with the fundamental plane, and their stellar velocity dispersions are $\sigma\approx 100-230$~km~s$^{-1}$ \citep[e.g.][]{genzel2001,veilleux2006}.
Furthermore, the comoving number density of ULIRGs at $z \approx  2$ is about three orders of magnitude larger than the local ULIRG density, consistent with most of the resulting early type galaxies being in place at redshift $z \approx 1.5$, and evolving further mostly by dry mergers \citep{lonsdale2006}. Local ULIRGs show evidence for extended thermal X-ray emission with $kT\approx 0.7$~keV, which has been interpreted as a result of galactic superwinds \citep{franceschini2003}. In systems such as Mrk~321 and IRAS~19254-7245 the X-ray emission is observed out to $r\approx 30$~kpc.  

\subsection{Observational Constraints on Hot Atmosphere Formation}

Hot atmospheres of giant elliptical galaxies likely formed from shock heated gas during early infall and from stellar ejecta. \citet{goulding2016} combined the high mass galaxies from the MASSIVE survey \citep{ma2014} with lower mass galaxies from the ATLAS$^{\rm 3D}$ survey \citep{kim2015,su2015} to study the X-ray and optical properties of a statistically significant sample of early-type galaxies. Their results show that while the thermalisation of stellar ejecta (mass loss, supernovae) is a significant source of hot gas \citep[see also][]{mathews2003,sun2007,sarzi2013}, the data are inconsistent with stellar mass loss alone. On the other hand, \citet{pellegrini2018} find that a major part of the observed $L_{\rm X}$ can be accounted for by the mass input from the stellar population.

\begin{figure}
\begin{center}
  \includegraphics[width=10cm]{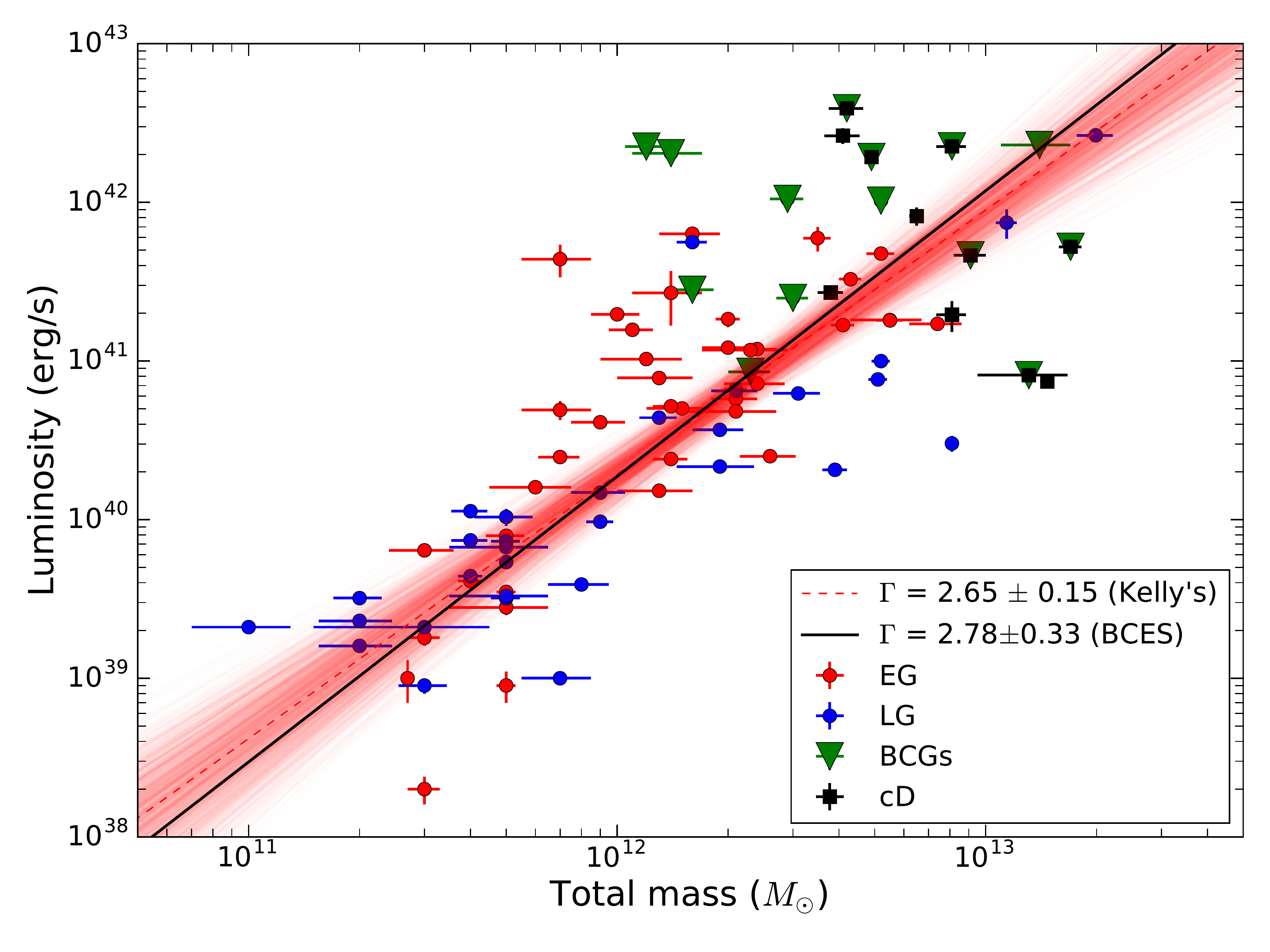}
 \end{center}
\caption{The X-ray luminosities of the hot atmospheres vs. the total mass within 5 effective radii for the early type galaxy sample of \citet{babyk2018a}.  ``EG" indicates elliptical galaxies; ``LG" indicates lenticular or S0 galaxies; ``BCGs" and ``cD" represent central cluster galaxies. The total gravitating masses were derived assuming the atmospheres are in hydrostatic equilibrium. The relation shows  large scatter, particularly at high-masses, most of which are central cD and brightest cluster galaxies. Some scatter may be attributed to effective radius measurements which vary widely between galaxies and observers.   }
\label{luminosity}       
\end{figure}

Halo mass appears to be the most important factor determining the X-ray luminosities of galaxies. \cite{forbes2017} used the total dynamical mass and X-ray gas luminosities of 29 massive early-type galaxies from the SLUGGS survey to probe $L_{\rm X}$-mass scaling relations \citep[see also][]{kim2013}. They found a strong relationship between $L_{\rm X}$ and galaxy dynamical mass within $5R_{\rm e}$, which is consistent with the cosmological simulations that incorporate mechanical heating from AGN \citep[e.g.][]{choi2015}. \citet{babyk2018a}, using the much larger sample shown in Fig.~\ref{luminosity}, found a similar correlation. They conclude that the hot gas was shock heated as it fell into collapsing dark matter halos so that $L_{\rm X}$ is primarily driven by the depth of a galaxy's potential well and heating by the AGN plays an important but secondary role in determining $L_{\rm X}$. The X-ray luminosities of the lower mass, cuspy, rotating galaxies do not correlate with their dynamical masses \citep{james2018}. In general, their atmospheric X-ray luminosities are low, indicating that they cannot effectively hold on to the hot gas. \citet{james2018} point out that the hot gas in these systems could be in an outflow state \citep[see][]{pellegrini2012}. Mass and energy input from stellar mass loss may be important in these galaxies.

Constraints on hot atmospheres have also been derived from measurements of the spectral distortions that hot atmospheres imprint on the microwave background.
\citet{spacek2016} used Sunyaev-Zeldovich (SZ) observations formed by stacking South Pole Telescope data of 3394 large, passive, galaxies in the redshift range $0.5 \leq z \leq 1.0$. Their average stellar mass is approximately $1.5\times 10^{11} M_\odot$. All are located outside of galaxy clusters. The measurement yielded a $\gtrsim 3 \sigma$ detection of the SZ effect on the scale of individual galaxy halos. The amplitude of the SZ distortion is directly proportional to the line-of-sight integral of the pressure. Therefore, the measurement has the potential to probe the average internal energy profile of distant, large elliptical galaxies. 

The  \citet{spacek2016} measurement suggests significant energy input to hot atmospheres from quasar-mode AGN feedback. However, a similar study using the Atacama Cosmology Telescope is consistent with models without significant energy input from AGN \citep{spacek2017}. While inconclusive, these studies indicate that ground-based measurements are approaching the sensitivity needed to place strong constraints on evolutionary models of hot atmospheres and their nuclear black holes \citep{chisari2018,spacek2018}. Future measurements with the next generation of ground-based microwave background telescopes should dramatically improve the situation.

\subsection{Hot Atmospheres and Stellar Rotation}

X-ray observations have shown that flattened systems, including massive spirals, S0s, and rapidly-rotating, coreless  ellipticals, have lower X-ray luminosities than round galaxies of the same optical luminosity $L_{\rm B}$ \citep{eskridge1995,sarzi2013}. Therefore, they are less studied and are thus poorly understood. Using the ATLAS$^{\rm 3D}$ integral field spectroscopy survey, \citet{sarzi2013} found that slow rotators, on average, have higher atmospheric temperatures and higher X-ray luminosities than fast rotators. The X-ray luminosities of fast rotators tend to decrease with the increasing degree of rotational support \citep[see also][]{boroson2011}.  

\citet{negri2014a,negri2014b} investigated the X-ray luminosities and temperatures of rotating galaxies using numerical simulations. They assumed that most of the atmospheric gas formed from thermalised stellar mass loss. They as well found systematically lower X-ray surface brightnesses and temperatures in rotating galaxies than in the non-rotating systems of the same mass. This can be understood as the combined effects of the centrifugal barrier of a rotating atmosphere and the decreased depths of the gravitational potential well owing to rotational support. Their simulations indicate that the relative importance of flattening and rotation on the X-ray luminosity and temperature of the hot atmospheres is a function of the galactic mass. While flattening and rotation in lower-mass galaxies promote winds that effectively lower their X-ray luminosities, mass loss from higher-mass, rotating, galaxies is negligible.

\subsection{The Great Divide: Galaxies and Missing Baryons}

Lower mass galaxies, those in halos significantly below $\approx 10^{12}~M_{\odot}$, are not expected to host hot atmospheres. Milky Way mass galaxies appear to be on the border between X-ray bright and X-ray faint systems. Deep X-ray observations of hot gas surrounding Milky Way sized elliptical, S0, and early spiral galaxies would place strong constraints on galaxy formation models. Many models indicate that as much as half of the warm-hot diffuse baryons in the local universe may lie in pressure-supported atmospheres around galaxies \citep[e.g.][]{fukugita1998,keres2005,fukugita2006}. Direct evidence for extended, volume-filling, atmospheres is scarce.

The best evidence for a hot atmosphere in the Milky Way is found from emission measurements of the soft X-ray background \citep{Snowden1997,Henley2010,Henley2012} and X-ray absorption measurements along sightliness to bright AGN \citep{paerels2003,Gupta2012}. \cite{Gupta2012}, combined {\it Chandra} observations of \ion{O}{vii} and \ion{O}{viii} absorption lines with {\it XMM-Newton} and {\it Suzaku} measurements of the emission measure of the Galactic component of the soft X-ray background, to find a hot atmospheric gas mass of over $6\times10^{10}~M_{\odot}$. 
\citet{Henley2012} developed an all-sky catalog of \ion{O}{vii} and \ion{O}{viii} emission line measurements. \cite{Miller2013,Miller2015} used  the catalog to determine an atmospheric gas mass of $\approx  10^{10} M_\odot.$  In their study of H$\alpha$ recombination emission from the Magellanic Stream, \citet{Tepper2015} developed a model of the hot Galactic atmosphere that was consistent with the observed stellar halo dynamics and with the UV/X-ray measurements, finding a mass of $2.5 \pm 1 \times 10^{10} M_\odot.$ These estimates show that the hot atmosphere contains a significant fraction of the baryonic mass, but still falls short of the $\approx  10^{11} M_\odot$ that would be needed to reach the universal value \citep{Bland-Hawthorn2016}.

X-ray atmospheres extending well beyond the stellar light have been observed only around nine massive, relaxed spiral galaxies \citep{anderson2011,dai2012,Bogdan2013a,bogdan2013b,li2017}. 
Using deep {\it XMM-Newton} observations of the two X-ray brightest spirals, NGC~6753 and NGC~1961, \citet{bogdan2017} and \citet{anderson2016}, respectively, find low metallicities of $Z\approx 0.1-0.2$ Solar in their hot atmospheres. They interpret the low metallicities as an indication that the hot gas was accreted from the surrounding environment. By extrapolating the density profiles of the galaxies to the virial radius, the authors estimate their total baryon mass fractions, finding that more than half of the baryons are missing.

\citet{li2018} stacked the X-ray data of six local isolated massive spiral galaxies from their Circum-Galactic Medium of MASsive Spirals (CGM-MASS) sample. They find that the mean gas density profile can be characterized by a single power law out to $r\approx  200$~kpc, which is about half the virial radius of the dark matter halo. They also find that more that 60--70\% of the baryons are missing within the virial radii of these spiral galaxies. These results imply that a significant fraction of the baryons escaped from their halos. 

Based on X-ray and Sunyaev-Zeldovich (SZ) observations, \citet{bregman2018} propose a consistent picture for the hot atmospheres of galaxies above approximately the Milky Way's luminosity. 
Stacked {\it Planck} measurements for massive, mostly early type galaxies reveal an SZ signal, which indicates that most of the baryons in these galaxies are hot and extend beyond their virial radii \citep{planck2013,greco2015}. However, the detected stacked SZ signal is nearly an order of magnitude larger than that inferred from the X-ray observations of massive spiral galaxies with $M_{\star} > 10^{11.3} M_{\odot}$. This result indicates, that there could be substantial differences in the hot gas atmospheres around massive spiral and early type galaxies. \citet{bregman2018} show that when the atmospheric density profiles of massive spirals are extrapolated to their virial radii, about half of their baryons are still missing from the hot phase. Only when extrapolated to $1.9-3~R_{200}$ does the baryon to dark matter ratio approach the cosmic value.

The low metallicities of $Z\approx 0.1-0.25$ Solar measured or assumed in these studies may be underestimated. Owing to pitfalls in modeling a multi-temperature gas with single temperature and omissions and incompleteness of atomic data tables for low temperature plasmas \citep[see][]{mernier2018}, the derived gas-phase metallicities are highly uncertain. \citet{Miller2015} have shown that the Milky Way's atmospheric metallicity must be higher than 0.3 Solar if it is to be consistent with pulsar dispersion measures toward the Large Magellanic Cloud. If the metallicities of the hot atmospheres of massive spirals are indeed strongly underestimated, then their true baryon fractions must be even lower than inferred. 


Several studies have focused on X-ray atmospheres of isolated elliptical galaxies, from fossil groups down to the mass of the Milky Way \citep{osullivan2004,osullivan2007,humphrey2011,humphrey2012b,humphrey2012}. For the fossil group RX J1159+5531, \citet{humphrey2012} combined {\it Chandra}, {\it Suzaku} and {\it ROSAT} data to study its X-ray emitting atmosphere out to $R_{100}$.  Within the virial radius, they measure a baryon fraction of $f_{\rm b} = 0.17\pm0.02$, consistent with the cosmological value. A similar result, a baryon fraction of $f_{\rm b}=0.13\pm0.3$ is obtained for the isolated $L_\star$ galaxy NGC 1521 \citep{humphrey2012b}. In the detailed study of the ``Milky Way sized'' NGC~720, \citet{humphrey2011} find that the baryon fraction within the virial radius is consistent with the cosmological value, confirming the theoretical predictions that an approximately Milky Way mass ($M_{\rm vir} = 3.1\pm0.4\times10^{12} M_\odot$) galaxy can sustain a massive hot atmosphere.  

Current observational constraints thus indicate, that while the hot X-ray emitting atmospheres surrounding Milky Way mass ellipticals are baryonically closed, spiral galaxies only hold on to a fraction of the hot baryons within their virial radii. 

\subsection{Chemical Constraints on the Galactic Atmospheres}

The metallicities of galaxy atmospheres encode, in principle, important information about their star-formation histories. If galaxies are baryonically closed, then the hot gas will contain the integrated yields of all supernovae and stellar winds accumulated throughout their lifetimes. If galactic feedback is strong enough to expel part of the hot baryons from the gravitational potential well of the galaxy, some of the metals produced in the period of maximum star-formation and galactic feedback activity might also be missing in modern atmospheres. 

Contrary to earlier results \citep[e.g.][]{rasmussen2009,bregman2010,sun2012,yates2017}, \citet{mernier2018} found that the cores of ellipticals, groups, and massive clusters, spanning two decades in mass from $\approx 10^{13}~M_\odot$ to $\approx 10^{15}~M_\odot$, have remarkably similar iron abundances within the radius $0.1r_{500}$. These systems have an O/Fe ratio  consistent with the Solar value to within $\approx  25\%$. This is true for clusters, groups, and giant elliptical galaxies \citep{deplaa2017}. The $\alpha$/Fe ratio in hot galactic atmospheres is thus different from the abundance ratios in the stellar populations of their host giant ellipticals, which typically have $\alpha$/Fe ratios twice as high \citep[e.g.][]{conroy2014}. This difference is consistent with the largely primordial origin of galactic atmospheres. The mass-invariance of the chemical enrichment of massive haloes provides additional support to the scenario of an early enrichment where the bulk of metals in hot atmospheres was in place well before clusters assembled \citep[see the reviews by][]{mernier2018c,biffi2018}. 

\section{Evidence for Galaxy Atmospheres Stabilized by Feedback}
\label{sec3}

Consensus has emerged that mechanical feedback from radio sources in central cluster galaxies suppresses cooling of hot, cluster-scale atmospheres. Clusters are relatively well understood because they can be imaged exquisitely in X-rays using tens of thousands, and in several instances, millions of photons. They are the only systems where cooling can be traced using a variety of thermodynamic diagnostics over eight decades of temperature and density.  Likewise, molecular gas levels and star formation rates in central brightest cluster galaxies (BCGs) often exceed those of gas-rich spirals, allowing a strong connection to be made between the cooling atmosphere, cold gas, and star formation.
Clusters are the standard against which other systems may be compared and understood.

\begin{figure}
\begin{center}
  \includegraphics[width=10cm]{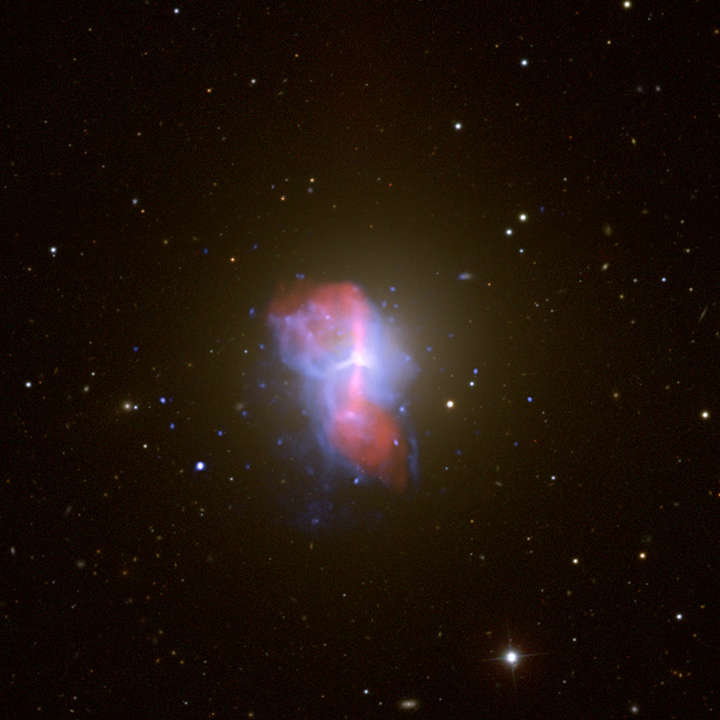}
 \end{center}
\caption{This composite image shows M~84, a massive elliptical galaxy in the Virgo Cluster. The {\it Chandra} image of the hot X-ray emitting gas is shown in blue and the VLA radio image of the jet injected relativistic plasma is shown in red. A background image from the Sloan Digital Sky Survey is shown in yellow and white. The jets inflated bubbles form cavities in the hot galactic atmosphere like those seen in clusters of galaxies. Credit: X-ray (NASA/CXC/MPE/A.Finoguenov et al.); Radio (NSF/NRAO/VLA/ESO/R.A.Laing et al); Optical (SDSS).}
\label{M84}       
\end{figure}

Like centrals in cooling clusters, the cooling times of giant elliptical atmospheres fall below tens of Myr within a kpc or so of the nucleus.  Radiation losses must be balanced by heating to avoid cooling at rates of 0.1--0.5 $M_{\odot}$~yr$^{-1}$ \citep{bregman2005} and star formation at comparable rates. The observed star formation rates in nearby, early type galaxies are generally below 0.1--0.01 $M_{\odot}$~yr$^{-1}$ \citep[e.g.][]{oconnel1999}. Using ultraviolet {\it Hubble Space Telescope} Wide Field Camera 3 imaging, \citet{ford2013} identified individual young stars and star clusters in four nearby giant ellipticals providing the best measurements of the star formation rates in these systems.  The rates are approximately $10^{-4}$--$10^{-5}~M_\odot$~yr$^{-1}$.  

As in cluster centrals, observations show that radio-loud AGN deposit enough enthalpy in the atmospheres of ellipticals to prevent cooling. But for the non-BCG giant ellipticals the relationship between hot atmospheres, cooling gas, and feedback is less clear.  The putative cooling and heating cycle is hard to observe, due largely to low X-ray photon count rates and low levels of warm and cold gas. Indications are that ellipticals are lower mass analogs to cluster centrals.
The total atmospheric gas masses of ellptical galaxies lie between $10^9 ~ M_\odot$ and $10^{10} ~ M_\odot$. 

\begin{figure}
\begin{center}
  \includegraphics[width=10cm]{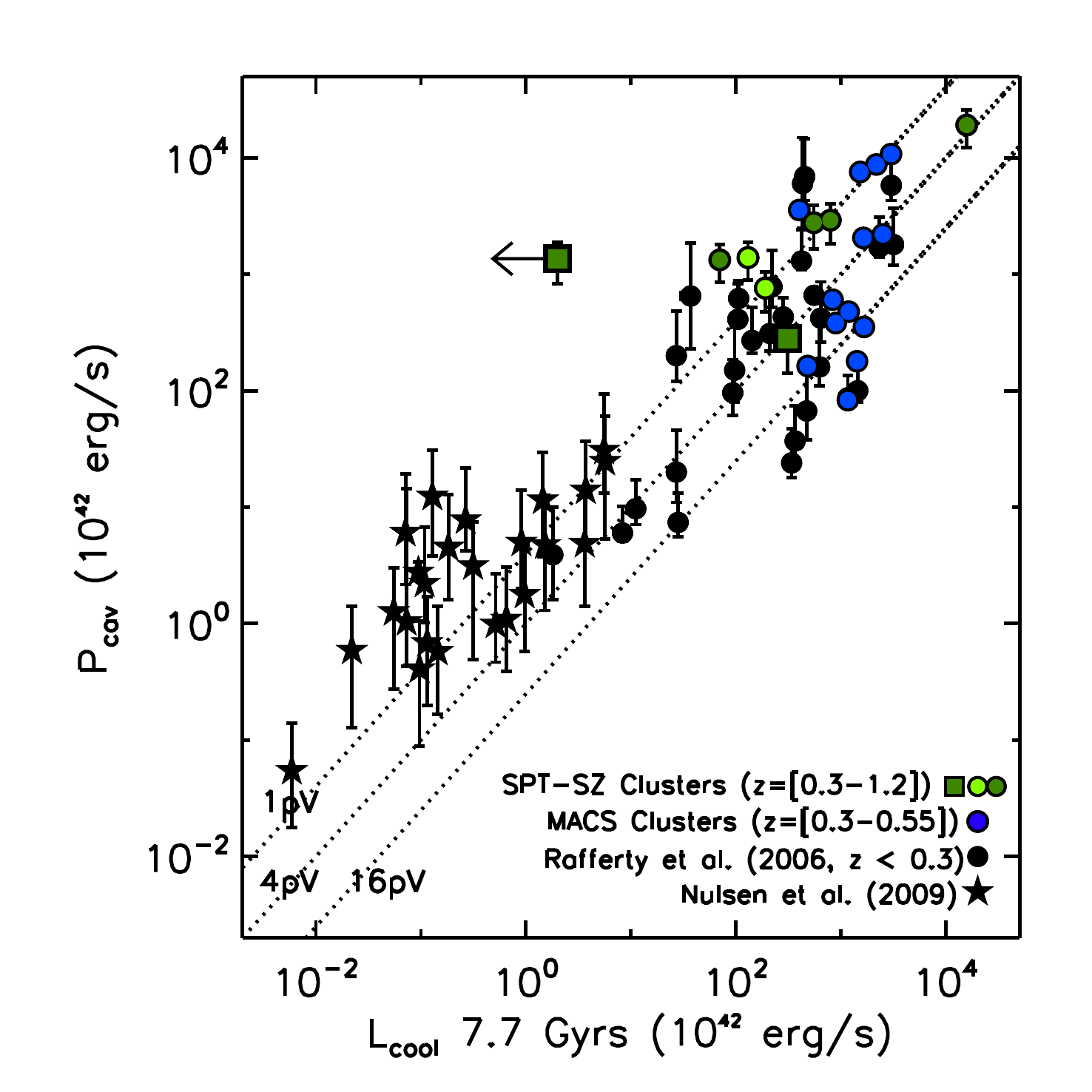}
 \end{center}
\caption{Mechanical power injected by the central AGN estimated from the energies and time-scales required to inflate the cavities observed in the surrounding hot atmospheres ($P_{\rm cav}$) versus the cooling luminosity within the cooling radius ($L_{\rm cool}$) from \citet{larrondo2015}. Dotted lines show $P_{\rm cav} = L_{\rm cool}$ for energy inputs of $1pV$, $4pV$ and $16pV$ per cavity, respectively, top to bottom. The observed relation indicates that the radio-mode feedback energy is sufficient to offset cooling in systems across six orders of magnitude in X-ray luminosity from the hot atmospheres of ellipticals to the most massive clusters of galaxies.}
\label{nulsen2009}       
\end{figure}

Early indications that giant elliptical galaxy hot atmospheres may be stabilized by feedback
were found in a {\it Chandra} X-ray observation of M~84 \citep{finoguenov2001b}. M~84 is a member of the Virgo cluster but does not reside in its centre.  Thus it harbours its own atmosphere. Its
radio jets and lobes inflated bubbles in its atmosphere (see Fig.~\ref{M84}) like those seen earlier in several cluster galaxies. The total enthalpy of M~84's X-ray cavities is $4pV=1.3\times10^{56}$ erg \citep{finoguenov2008}, which is far lower than the values estimated in clusters. Nevertheless this figure is energetically significant compared to the thermal energy of the atmosphere.

A spectacular example is the sequence of three bubble pairs observed by \citet{randall2011,randall2015} in the group central galaxy NGC~5813. The pairs of bubbles are aligned in radius lying at projected
altitudes of 1 kpc, 8 kpc, and 20 kpc from the nucleus. They were launched 3 Myr, 20 Myr, and 90 Myr ago, respectively. The bubbles are surrounded by shock fronts with Mach numbers between 1.5 and 1.7.  The total energy released exceeds several $\times 10^{57}$ erg, which is sufficient to offset the cooling of the atmosphere.  
\citet{nulsen2009} examined a sample of 24 ellipticals with cavities from the {\it Chandra} archive to determine whether they are able to systematically offset cooling of the atmospheres. Plotting the jet power,
determined from the enthalpy of the X-ray cavities, against the X-ray luminosity of the cooling atmosphere, \citet{nulsen2009} concluded that radio/mechanical feedback energy is sufficient to offset cooling. Similar results were found in cluster centrals \citep{birzan2004,rafferty2006,larrondo2015}, and in a large, statistical sample of Sloan ellipticals \citep{Best2006}. This continuity between clusters and giant ellipticals, shown in Fig. \ref{nulsen2009}, is a promising sign that elliptical atmospheres are likewise stabilized by radio mechanical feedback.

\subsection{Cool Gas in Elliptical Galaxies}

Elliptical galaxies were once thought to be free of gas. Lacking prominent gas disks and active star formation, they must have depleted their stores of cold gas during their formation epoch, billions of years in the past. That view changed with discoveries of dust lanes, nuclear gas disks, and nebular emission in many ellipticals \citep{sadler1985,goudfrooij1994,macchetto1996}. These features indicate low levels of cold molecular and atomic gas in ellipticals and S0 galaxies \citep{knapp1985,knapp1989,knapp1996}. 

\begin{figure}
\begin{center}
\hspace{-0.5cm}  \includegraphics[width=7cm]{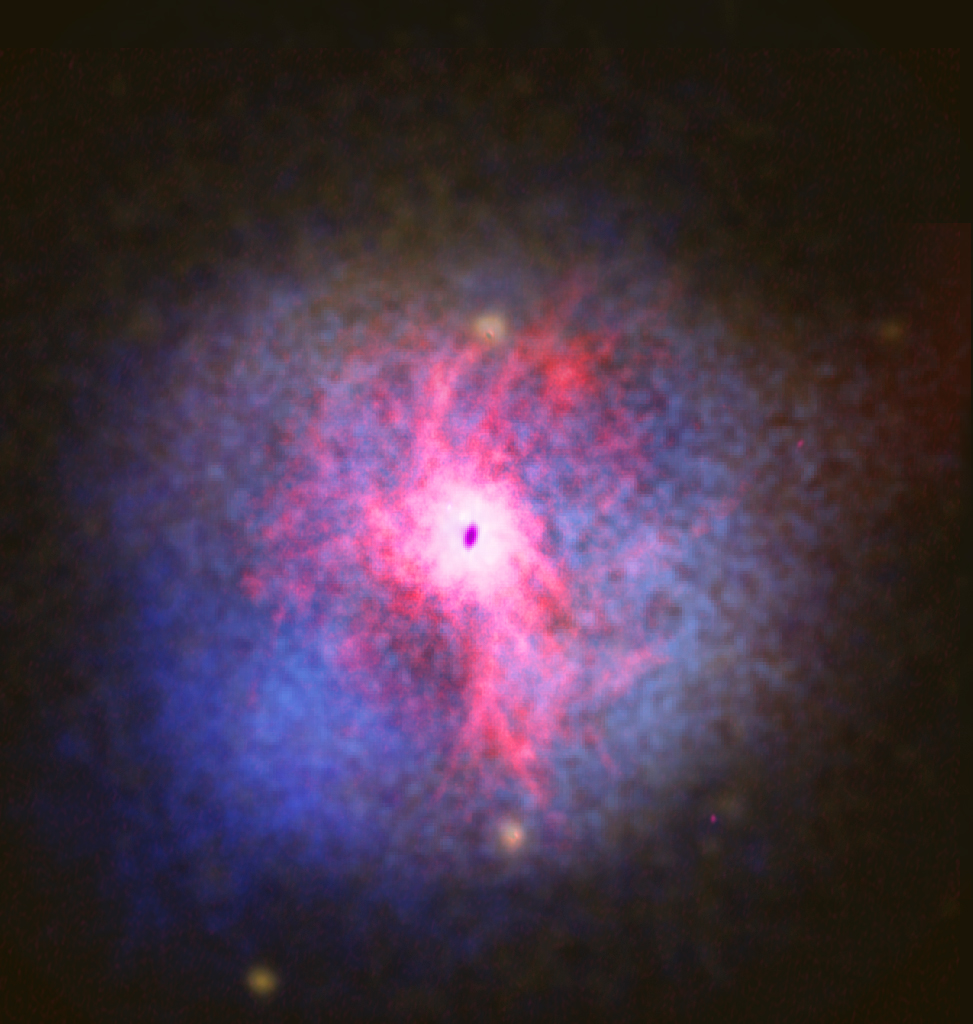}
\end{center}
\caption{Multi-wavelength, optical, narrow-band H$\alpha$ + [\ion{N}{ii}], X-ray, and radio composite view of the giant elliptical galaxy NGC~5044.
The stellar component, as observed at optical wavelengths (Digitised Sky Survey), is shown in white. 
The galaxy is embedded in a hot X-ray emitting atmosphere of ionised gas, which is shown in blue ({\it Chandra}). 
Observations show that some of the hot gas cools  and flows towards the centre of the galaxy. The filamentary network shown in red is warm H$\alpha$ + [\ion{N}{ii}] emitting gas observed by the Southern Observatory for Astrophysical Research (SOAR) telescope in Chile. When observed in radio at 1.4 GHz (VLA), this galaxy appears only as a point source shown in violet at the centre of the image. Credit: Digitised Sky Survey/Chandra X-ray Center/Southern Observatory for Astrophysical Research/Very Large Array \citep{werner2014}.}
\label{N5044}       
\end{figure}

Elliptical galaxies are rich in gas, but most of it resides in hot atmospheres (see Fig. \ref{N5044}) with masses of  $10^9 ~ M_\odot$ to $10^{10} ~M_\odot$ \citep{forman1985}.  Their molecular gas masses lie between $10^5 - 10^8 ~ M_\odot$. Therefore, atmospheric masses exceed cold gas masses by 100 times or more, but with large variance \citep{babyk2019}.  

\begin{figure}
\begin{center}
\hspace{-0.5cm}  \includegraphics[width=10cm]{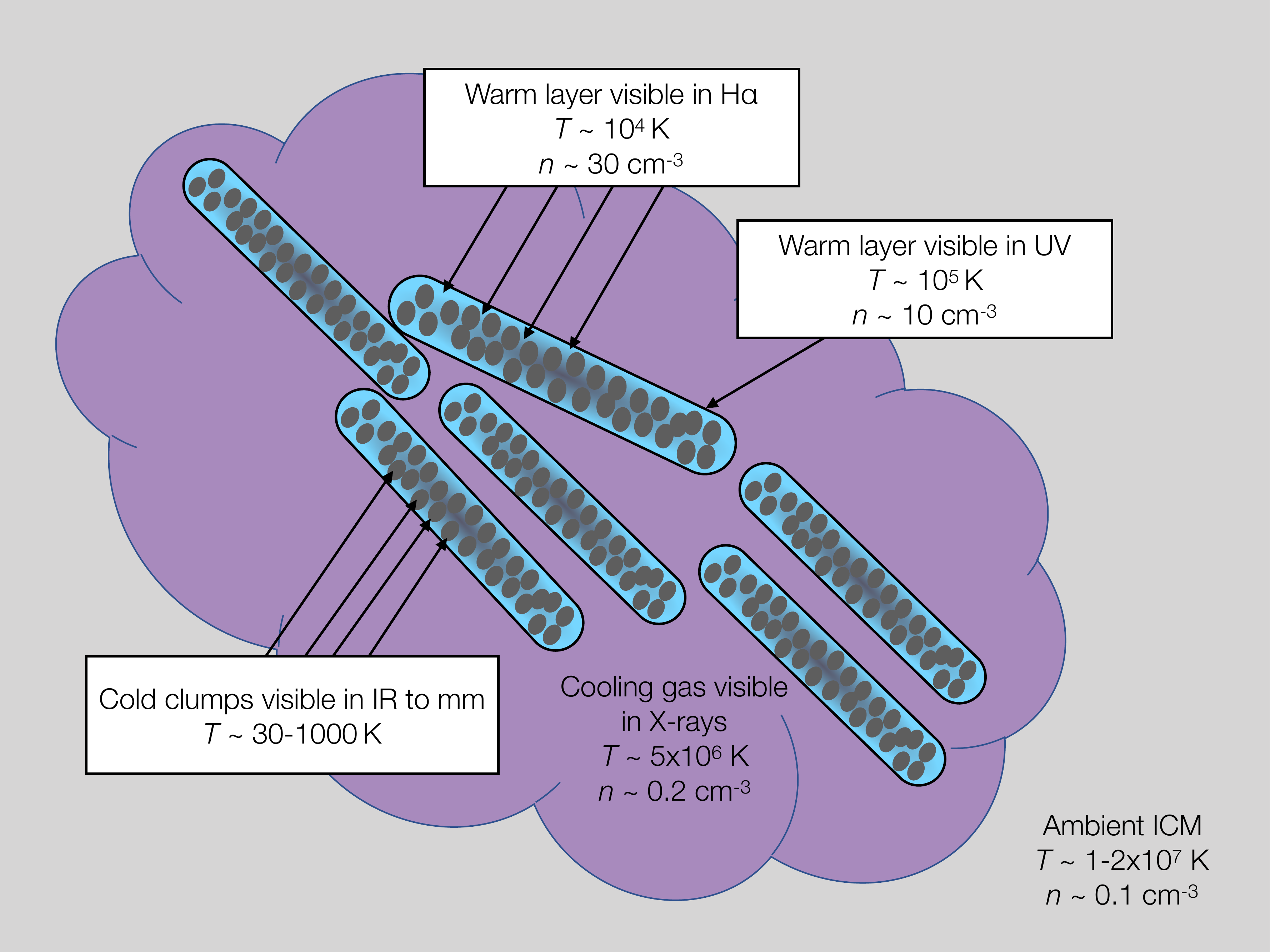}
\end{center}
\caption{A schematic view of the possible structure of the filamentary cool gas found in giant ellipticals. The filaments are composed of clumps of gas with low volume filling fractions with temperatures spanning from $\sim30$~K CO emitting molecular gas, through $\sim100$~K diffuse atomic [\ion{C}{ii}] emitting gas, to a warm $\sim1000$~K phase. They are enclosed within layers of warm $10^4$~K H$\alpha$+[\ion{N}{ii}] and $10^5$~K far-UV emitting gas phases. The filaments appear to be strongly magnetized with fields of several tents of $\mu$G. They are surrounded by cooling X-ray emitting gas, with a temperature of $\sim5\times10^6$~K, that is significantly cooler than the ambient $kT\sim2$~keV ICM \citep[based on][]{werner2013,anderson2018}.  }
\label{filaments}       
\end{figure}

About half of optical and X-ray bright giant elliptical galaxies display nuclear or extended filamentary H$\alpha$+[\ion{N}{ii}] emission \citep{lakhchaura2018}. Similar to the nebulae observed in the centres of cooling core clusters \citep{johnstone1987,heckman1989}, the line emission in these systems cannot be powered by the radiation of young, hot stars or the central AGN.  The line emission is too luminous to be recombination radiation from cooling. The filaments likely consist of strands with small volume filling fractions \citep{fabian2008,werner2013}. The soft X-ray emission associated with filaments in cool-core clusters \citep[e.g.][]{sanders2007,werner2010,werner2013} is cooling hot plasma enveloping the cooler threads (see Fig.~\ref{filaments}). The hot plasma may be cooling both radiatively and by mixing with the cold gas in the filaments \citep{sanders2009b,werner2013}. In M~87, the emission-line nebulae spatially coincide with \ion{C}{iv} line emission at far-ultraviolet (FUV) wavelengths.  This emission emerges from gas at temperature $\sim 10^5$ K indicating efficient energy transport between gas phases \citep{sparks2009,sparks2012}, possibly by electron thermal conduction or mixing.  This energy transport links the cool, optically-emitting filaments to the surrounding hot gas \citep[see also][]{anderson2018}. Mixing can in principle supply the power and the ionizing particles needed to explain the observed line emission \citep{ferland2008,ferland2009,fabian2011}. Depending on the mixing rate, this process can cool the ICM non-radiatively, or it may be evaporating the filaments. \citet{ferland2009} argue that the emission-line spectra of the filamentary nebulae around central galaxies of cooling core clusters most likely originate in gas exposed to ionizing particles, either relativistic cosmic rays or hot X-ray emitting plasma penetrating into the cold gas \citep[see also][]{canning2016}. 

Reconnection of magnetic fields in the wakes of buoyant bubbles has also been proposed as a mechanism for powering the filaments \citep{churazov2013}. 
The [\ion{S}{ii}]$\lambda$6716/6731 line ratios in filaments indicate low densities in the 10,000~K nebulae, based on which \citet{werner2013} argued that significant magnetic pressure (fields of several tens of $\mu$G) would be required to keep the warm ionised gas in pressure ballance with the surrounding intracluster medium (ICM). Significant magnetic fields threading the filaments were inferred using arguments based on the integrity of the filaments in the Perseus cluster \citep{fabian2008}, and based on radio observations of the Faraday rotation measure in cooling core clusters \citep{feretti1999,allen2001,taylor2001,taylor2007}. Magnetic support may also counteract gravitational collapse of gas clouds preventing them from forming stars.

In the sample of \citet{werner2014} all galaxies with extended optical emission nebulae contain cool ($\approx $100 K) atomic [\ion{C}{ii}] emitting gas.
The MIR {\it Spitzer} spectra of galaxies with emission nebulae also show the presence of dust, warm H$_{2}$ molecular gas and  polycyclic aromatic hydrocarbon (PAH) emission \citep{temi2007,temi2007b,panuzzo2011}.

The ATLAS$^{\rm 3D}$ volume-limited survey of CO line emission revealed cold molecular clouds in about one fifth of early-type galaxies \citep{young2011}. The best example of strong CO line emission observed in a nearby massive giant elliptical is NGC~5044. Using ALMA observations, \citet{david2014} detected many CO(2-1) emitting molecular clouds in the central region of the galaxy. These structures are likely giant molecular associations composed of individual molecular clouds with a volume filling fraction of about 15\%. Their masses are in the range between $3\times10^5~M_\odot$ to $10^7~M_\odot$. Given their CO(2-1) line widths, these structures are short-lived and they will disperse in about 12 Myr.

The total molecular gas mass in NGC 5044 is $\approx 6\times10^7~M_\odot$ \citep{temi2018}. Other group central giant elliptical galaxies have less molecular gas.
ALMA CO(2-1) observations of NGC 5846 and NGC 4636 revealed only $2\times10^6~M_\odot$ and $2.6\times10^5~M_\odot$ of molecular gas, respectively. The molecular gas mass associated with the emission line nebulae East of the nucleus of M~87 is also only $M_{\rm H_2} = (4.7\pm0.4)\times10^5 M_{\odot}$ \citep{simionescu2018}. These molecular clouds are apparently short lived, transient phenomena. 

Given the relatively large H$\alpha$+[\ion{N}{ii}] and [\ion{C}{ii}] luminosities of some of these galaxies, the inferred small molecular gas masses might seem surprising. If, however, the density of the molecular gas is low due to turbulence and collisions with hot gas or cosmic rays, then the CO(2-1) line emission will be suppressed, causing the molecular gas mass to be underestimated \citep{canning2016}. The additional pressure support in the cold gas may be slowing or preventing the gravitational collapse of any molecular gas clouds that exceed the Jeans mass, preventing star formation in these red and dead galaxies. 

Some of the cold clouds might get accreted onto the central supermassive black hole in NGC~5044, as suggested by the absorption features in the spectrum of the central continuum source, which reveal infalling clouds with a radial velocity of 250 km~s$^{-1}$ \citep{david2014}. Infall of molecular clouds was also inferred from the absorption spectra of the nucleus of the brightest cluster galaxy in Abell 2597 by \citet{tremblay2016}. 

Radio power correlates with total molecular gas mass in galaxies from low-power ellipticals and S0s to the most powerful central cluster galaxies \citep{babyk2019}.  This correlation is consistent with molecular fueling of radio galaxies. However, the scatter in the trend is large, approximately three decades in molecular gas mass at a given radio power. This scatter likely reflects conditions on large scales and not those nearest to the nucleus that are driving current radio activity \citep{mcnamara2011}.
Systems with low molecular gas masses may host powerful AGN and conversely so.

Cold gas is also observed in surveys of more distant early type galaxies. Several authors have studied cold gas around $z\approx 0.5$ Luminous Red Galaxies (LRGs) selected from the Sloan Digital Sky Survey using QSO absorption line techniques. These galaxies have colors similar to nearby ellipticals \citep{eisenstein2001}, stellar masses of $\gtrsim 10^{11} M_\odot$ \citep{tojeiro2011}, and clustering amplitudes consistent with halos masses $\gtrsim 10^{13} M_\odot$ \citep[e.g.][]{zheng2007}. \cite{gauthier2009} showed that the cross-correlation function of
LRG and \ion{Mg}{ii} absorbers  is similar to the LRG auto-correlation function at large separations.  This suggests
that the halos of a large fraction of these galaxies contain strong \ion{Mg}{ii} absorbers.
The absorbers trace photo-ionized, $T \approx 10^4$ K gas  \citep{bergeron1986} with
neutral hydrogen column densities $10^{16}-10^{22} \, {\rm cm}^{-2}$ \citep{rao2006}.
Other studies found similar
absorbers in a significant fraction of LRG halos 
at projected distances $ d \approx 100-500$ kpc
\citep{gauthier2010,huang2016}.  
Furthermore, strong Ly$\alpha$ absorption is common around $z\approx 0.2$  early-type galaxies \citep{thom2012,tumlinson2013},
but at lower levels than in star forming galaxies \citep{borthakur2016}.
However, using multiply-lensed QSOs, \cite{zahedy2017} found high column density gas around
$z=0.4-0.7$ ellipticals at projected distances as small as
$\approx 3-15$ kpc.  

Recently, \cite{Chen2018} combined Cosmic Origins Spectrograph absorption-line measurements with ground based measurements for a sample of 16 LRGs. They found an overall median \ion{H}{i} column density of $10^{16.6} \, {\rm cm}^{-2}$ and a mean \ion{C}{iii} covering fraction of $\approx 0.75$ for strong \ion{C}{iii} absorbers within a distance $\approx 160$ kpc.  These values are similar to those seen around typical $L_*$ galaxies. Apparently, massive quiescent halos contain widespread
chemically-enriched cool gas at a similar level to halos hosting significant star formation.

\subsection{Observations Indicating Condensation from Hot Atmospheres}

Whether the cool interstellar media observed as H$\alpha$+[\ion{N}{ii}] and CO emission in early type galaxies condensed from their hot atmospheres or were
accreted externally is unclear. That gas is accreted by mergers is indisputable. However, recent studies indicate that some cold gas, perhaps most, cooled from hot atmospheres.

Studying X-ray, far-infrared, and optical data for a sample of ten nearby giant ellipticals, \citet{werner2014} found that the galaxies with extended H$\alpha$+[\ion{N}{ii}] and [\ion{C}{ii}] line emission have lower atmospheric gas entropies and cooling times beyond 1 kpc than the cool gas-poor systems. This indicates that cool gas is related to and perhaps cooled from the hot atmospheres.
\begin{figure*}
\centering
  \includegraphics[width=\linewidth]{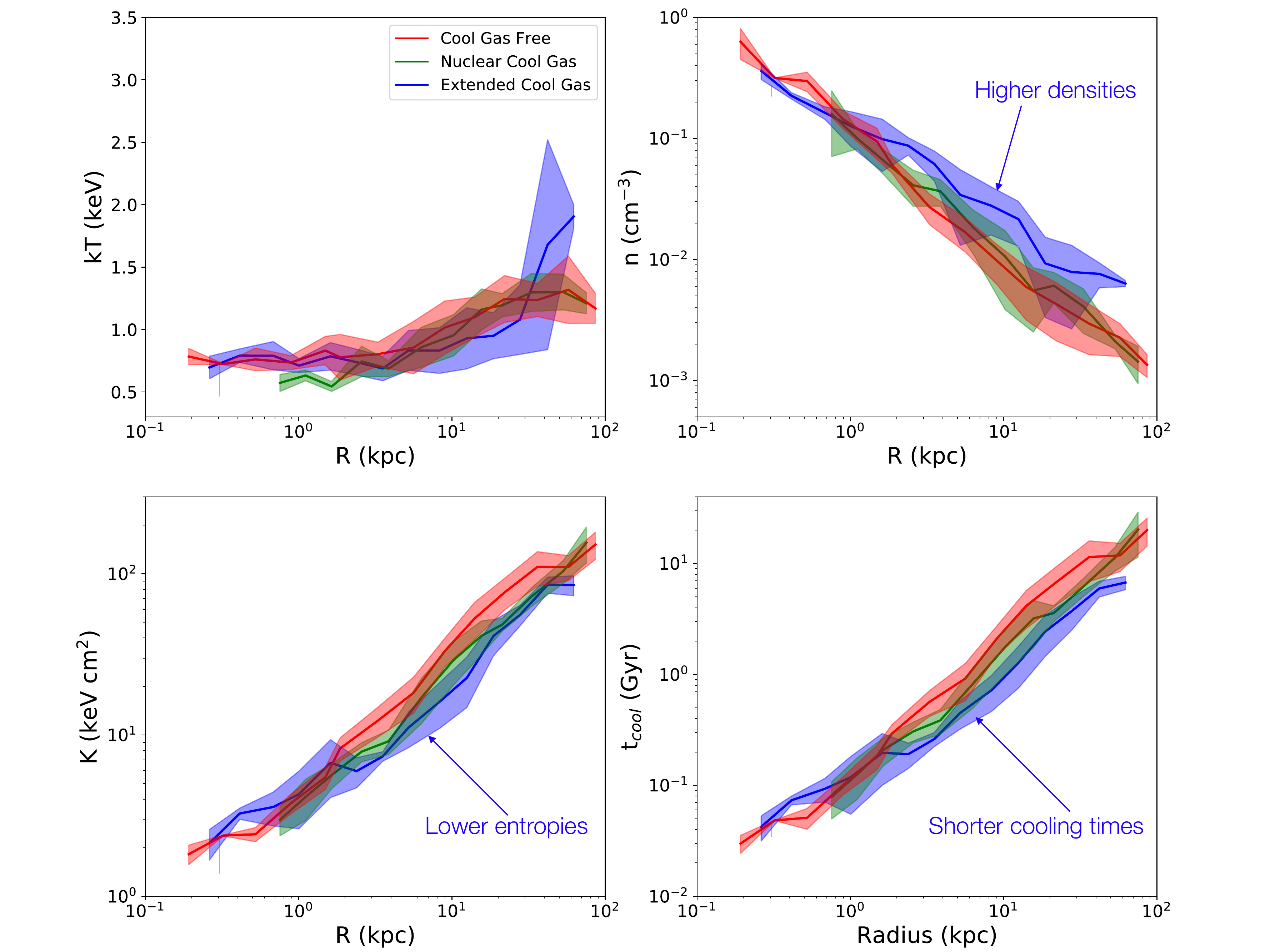}
\caption{Left: The combined radially binned profiles of temperature (top left), density (top right), entropy (bottom left) and cooling time (bottom right) 
for the sample of \citet{lakhchaura2018}.  The red, green and blue 
solid lines show median profiles for  the cool gas free, nuclear cool gas and extended cool gas systems, respectively and the shaded regions show the median absolute deviation (MAD) spreads about the medians. The figure shows higher densities and lower entropies and cooling times for the extended cool gas galaxies 
than the rest of the sample, outside the innermost regions ($\approx $2 kpc), albeit with significant spread. In the central $\approx 1$~kpc the mean  profiles might be affected by the lower resolution of some of the datasets.}
\label{fig:lakhchaura}
\end{figure*}
\citet{lakhchaura2018} expanded the sample to 49 nearby ellipticals observed with the {\it Chandra} X-ray Observatory.  Combining X-ray observations with H$\alpha$ + [\ion{N}{ii}] imaging, they found no correlation between nebular emission, and X-ray luminosity, atmospheric gas mass, or gas mass fraction.  However, nebular emission was more likely to be found in systems with the highest gas densities, lowest entropies, shortest cooling times, and shallower entropy profiles (see Fig. \ref{fig:lakhchaura}). Moreover, nebular emission is more likely to be associated with disturbed X-ray morphologies. X-ray disturbances may be linked to several factors, including mergers and AGN radio jets. The correlations found in this study are at the $\approx  2.5 \sigma$ confidence level. Nevertheless, they have been confirmed by independent X-ray and CO observations \citep{babyk2019}.

\citet{babyk2019} studied X-ray emission from 40 nearby ellipticals and early spirals observed by {\it Chandra}, a sample with strong overlap with Lakhchaura's. Babyk et al. examined the atmospheric properties of systems with single dish CO observations from the ATLAS$^{\rm{3D}}$ complete sample of nearby galaxies \citep{young2011}.  Molecular gas masses lie between $\sim  10^6 ~\rm M_\odot$ and $\sim  10^8 ~\rm M_\odot$. Their results are similar to \citet{lakhchaura2018}: all systems with CO detections  have atmospheric cooling times shorter than 1~Gyr at $R\approx 10$ kpc, while those with restrictive CO upper limits have longer atmospheric cooling times.   

\begin{figure}
\begin{center}
  \vspace{-3cm}
  \includegraphics[width=10cm]{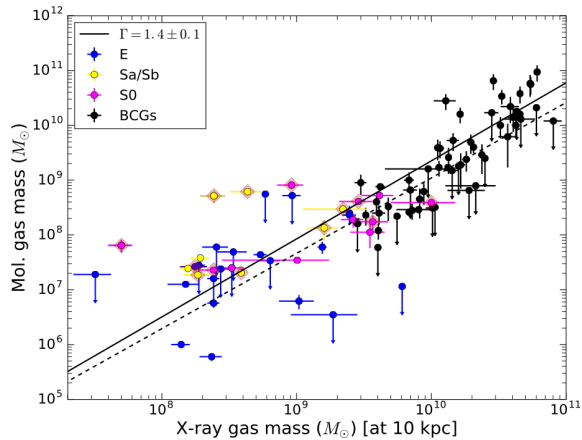}
  \vspace{-3.5cm}
 \end{center}
\caption{Molecular gas mass derived using CO observations \citep{edge2002,young2011,pulido2017} vs. hot X-ray emitting gas mass within the radius of 10 kpc for the early type galaxy sample of \citet{babyk2019}. The scatter increases at low masses, where the CO emission is weak and the measurements are noisy. The positive trend suggests that molecular gas is related to and perhaps cooled from the hot atmospheres.  }
\label{molecular}       
\end{figure}

Similar results were found for hot atmospheres surrounding central galaxies in galaxy clusters. \citet{rafferty2008} and \citet{cavagnolo2008} have shown that star formation, H$\alpha$ emission, and radio loudness ensue when the central cooling time and atmospheric entropy fall below approximately $10^9$ years and
about 30 keV cm$^2$, respectively. The sharp cooling time threshold at $\approx 10^9$ years ties radio/mechanical feedback to cooling hot atmospheres.

Finally, the mass of molecular gas in ellipticals and central cluster galaxies is apparently related to both the mass and cooling time of their hot atmospheres. \citet{babyk2019} combined the ellipticals from ATLAS$^{\rm{3D}}$ with the study of central cluster galaxies by \citet{pulido2017}. Figure \ref{molecular} shows a trend between the molecular gas mass and atmospheric mass within the radius of 10 kpc. The scatter is large and apparently increases in the ellipticals and early spirals.  Some scatter is related to the early spirals where molecular gas is located in disks.  These systems contain elevated levels of molecular gas for a given atmospheric mass.  The overall trend suggests that molecular gas and atmospheric gas are related to each other. In central cluster galaxies this is almost certainly due to atmospheric cooling. This also appears to be true for early type galaxies.

\subsection{Thermally Unstable Cooling}

The thermal stability of hot atmospheres is a topic of great importance to the evolution of clusters and galaxies. Hot atmospheres with short central cooling times were long ago shown to be thermally unstable   \citep{fabian1977,mathews1978}, and likely condensing from overdense gas parcels \citep{nulsen1986}. 
Others argued that small perturbations in a stably stratified atmospheres are thermally stable \citep[][]{malagoli1987,balbus1989}.
\citet{pizzolato2005} have shown that thermally unstable cooling should proceed in atmospheres from preexisting density perturbations.  The perturbations are assumed to be non-linear, seeded by the passage of an earlier radio jet. The density perturbations cool into molecular clouds feeding star formation and the nuclear black hole. 

Later work by  \citet{mccourt2011} showed that in a plane-parallel atmosphere, cooling parcels of gas are stabilized by buoyancy.  Thermally unstable cooling ensues from linear perturbations when the local value of the ratio of thermal instability timescale to the local free-fall timescale approaches unity. This condition is never satisfied in the atmospheres of clusters \citep{hogan2017} or elliptical galaxies \citep{babyk2018,lakhchaura2018}, where the $t_{\rm cool}/t_{\rm ff}$ ratios lie well above unity. \citet{mccourt2012} and \citet{sharma2012} went on to suggest that in a realistic, spherical atmosphere, thermally unstable cooling would proceed from low-amplitude, linear density perturbations when the ratio of the cooling time to the free-fall timescales falls below 10. This potentially exciting result opened the door to subsequent studies that likewise suggested that thermally unstable cooling ensues in atmospheres when $t_{\rm cool}/t_{\rm ff} \lesssim 10$ \citep{gaspari2012,gaspari2013,voit2015a,voit2015b,voit2015c,voit2015d,prasad2015}.

The conjecture that thermally unstable cooling ensues from linear density perturbations when $t_{\rm cool}/t_{\rm ff}\lesssim10$ was explored in several studies \citep{mcnamara2016,hogan2017,pulido2017}.  The most extensive test was conducted on a large sample of clusters by \citet{hogan2017}. Close attention was paid to resolution effects \citep{panagoulia2014}, and to the difficult problem of measuring mass profiles into the cores of clusters \citep{hogan2017a}. They found that the ratio of $t_{\rm cool}/t_{\rm ff}$ rarely if ever falls below 10.  Systems with the signatures of thermally unstable cooling, including star formation and molecular clouds, have central $t_{\rm cool}/t_{\rm ff}$ values lying between 10 and 35.  These values lie well above the classical threshold $t_{\rm cool}/t_{\rm ff} < 1$ and significantly above 10. No correlations between the minimum value of $t_{\rm cool}/t_{\rm ff}$ and molecular gas mass \citep{pulido2017}, H$\alpha$ emission \citep{cavagnolo2008}, or star formation \citep{hogan2017,prasad2017} were found. Only a threshold is observed: systems contain nebular emission, molecular gas, and star formation when the central cooling time falls below $10^9$ yr. These conclusions also follow for giant elliptical and early spiral atmospheres in \citet{babyk2018}, \citet{lakhchaura2018}, and \citet{babyk2019}.

The theoretical motivation for the $t_{\rm cool}/t_{\rm ff}\lesssim10$ threshold is unclear. \citet{sharma2012} suggested that in realistic, three dimensional atmospheres the threshold for thermally unstable cooling rises upward from unity by an order of magnitude to a value of 10.  However,  the more recent study by \citet{choudhury2016} did not confirm this conjecture.  

That is not to say that the ratio $t_{\rm cool}/t_{\rm ff}$ is irrelevant. Observation shows that the cooling time is driving the ratio. 
While mass profiles of central cluster galaxies are remarkably similar, the cooling time profiles are not.
Central values of atmospheric cooling time span a decade, but the free-fall times lie within a factor of 2 of each other \citep{hogan2017}.
Systems with lower values of $t_{\rm cool}/t_{\rm ff}$ have shorter atmospheric cooling times and are thus prone to thermally unstable cooling.  Nevertheless, the farther from unity this ratio lies, the longer the cooling time, and thus the more thermally stable the atmosphere. 

How do you satisfy $t_{\rm cool}/t_{\rm ff}\lesssim 1$ in atmospheres where the average radial value always exceeds 10? Two mechanisms have been proposed: 1) lifting gas in the updrafts of rising radio bubbles \citep{mcnamara2016}, and atmospheric turbulence \citep{voit2018,gaspari2018}. Both approaches amount to much the same thing: cooling parcels of gas near the nucleus of the galaxy are lifted to higher altitudes where the free-fall time, or more realistically the infall time \citep[see][]{mcnamara2016}, is increased until the ratio approaches unity.  In the first instance gas is lifted directly. In the second it is lifted and circulated in turbulent eddies. 

X-ray and sub-millimeter observations have connected uplift behind rising bubbles to the production and/or displacement of molecular clouds in BCGs \citep{salome2011,russell2017}. The situation in central group galaxies and non-BCG giant ellipticals is less clear but likely the same process is occurring there \citep[see Fig. \ref{N5044};][]{david2014,werner2014,randall2015}. Any dynamical event, such as a merger or gas sloshing, that displaces thermally unstable gas may stimulate cooling through direct uplift or by generating turbulence. 

The question then becomes: which mechanism is more efficient? Turbulence must be generated and sustained by an action. The most important sustained action is AGN feedback \citep{voit2018,gaspari2018}, which also drives uplift. Therefore, it is not clear whether turbulence would be the primary mechanism driving thermally unstable cooling. Turbulence would probably enhance the effects of direct uplift, which has been shown by observation to be remarkably efficient. On the other hand, turbulence at some level will be present in all atmospheres.  Understanding its relationship to thermally unstable cooling is a new and potentially important theoretical development.

The minimum value of the $t_{\rm cool}/t_{\rm ff}$ profile may be less important than its shape.  The mass profiles of the inner regions of central galaxies are best modeled as isothermal spheres \citep{churazov2010,lyskova2014,hogan2017a}. Their entropy profiles are likewise remarkably similar, with $K\propto R^{2/3}$ power law shape in the inner $10-20$ kpc or so, and $K\propto R^{1.1}$ in the atmosphere beyond \citep{panagoulia2014}. The $K\propto R^{2/3}$ form is a general property of the atmospheres of centrals in clusters, giant ellipticals and early spirals \citep[][see also Sect.~\ref{similarity}]{babyk2018a}.  \citet{hogan2017a} pointed out that an isothermal potential and a flat (constant) $t_{\rm cool}/t_{\rm ff}$ profile naturally leads to the $K\propto R^{2/3}$ form.  Indeed, in the few systems where $t_{\rm cool}/t_{\rm ff}$ profiles are resolved in their centers they are flat or nearly so. It is in this region where molecular clouds and star formation are observed and where the atmosphere is prone to thermally unstable cooling. It seems that feedback, likely mechanical AGN feedback, stabilizes atmospheres in a way that leads to these general thermodynamic profiles.  This statement may be tested in the future using numerical simulations.

\subsection{The Thermodynamic Similarity of Galactic Atmospheres}
\label{similarity}

\begin{figure}
\begin{center}
\hspace{-0.5cm}  \includegraphics[width=10cm]{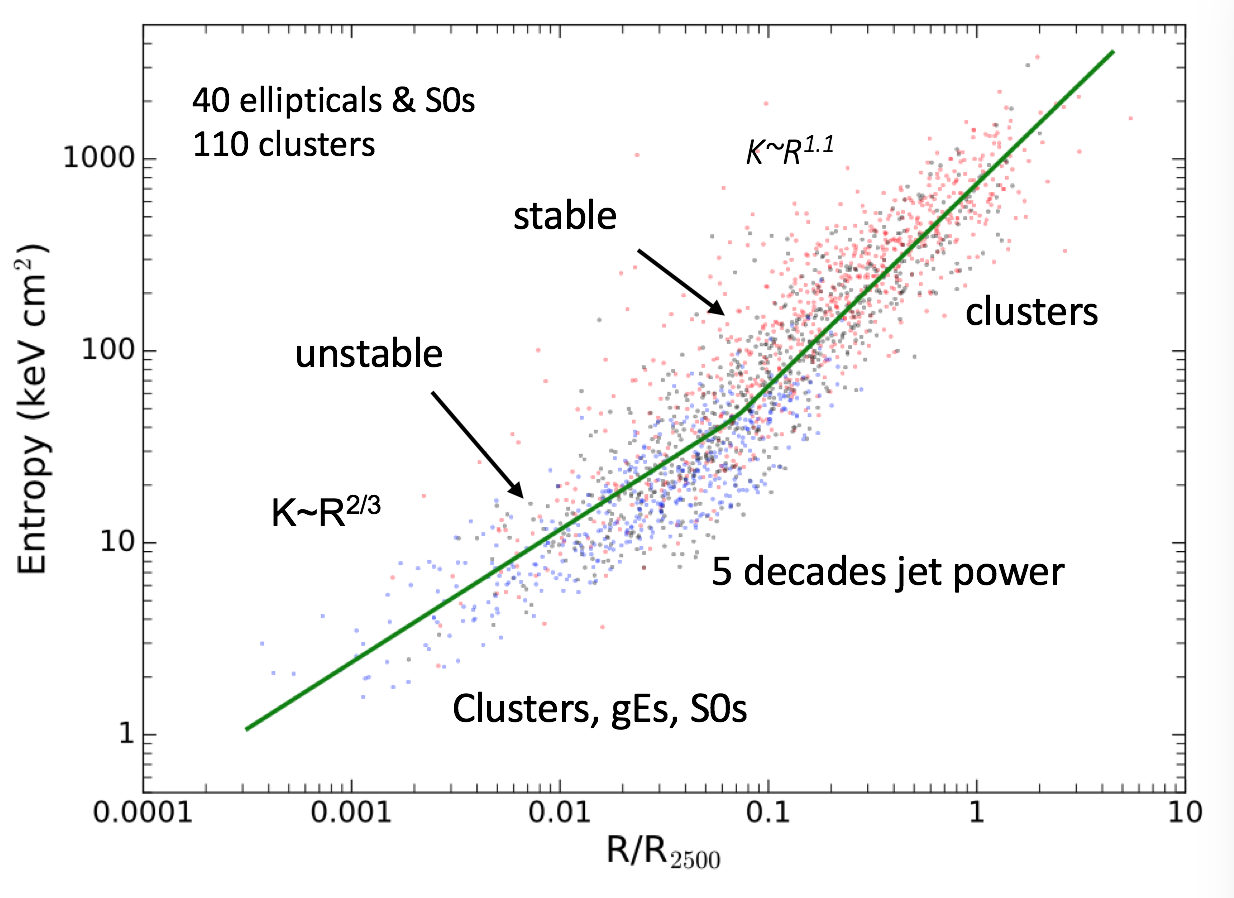}
\end{center}
\caption{The deprojected radial entropy profiles fitted by a broken power law model from \citet{babyk2018}.
The error bars were omitted for clarity. Beyond approximately $0.1R_{2500}$ the entropy is distributed as $R^{1.1}$ and the atmosphere is thermally stable, while at smaller radii the entropy distribution is flatter, $R^{0.67}$, and the gas is prone to thermal instabilities.}
\label{babyk}       
\end{figure}

The central thermodynamic profiles of the hot atmospheres of galaxies, groups, and clusters of galaxies show a remarkable similarity. Notably, the entropy profiles follow a form $K\propto r^\alpha$ with $\alpha$ in the range $\approx  0.5-1.1$ between 1--30 kpc \citep{babyk2018,lakhchaura2018}.  \citet{babyk2018} show that the entropy profiles of all systems can be fit with broken power-laws with $\alpha\approx 2/3$ within $\approx 0.1R_{2500}$ and  $\alpha\approx 1$ at larger radii (see Fig.~\ref{babyk}). The gas entropy in the outer parts of hot atmospheres is determined primarily by gravitational collapse and follows $K\propto r^{1.1}$ \citep{tozzi2001,voit2005}. The entropy in the vicinity of the central galaxy at $r\lesssim15$ kpc is significantly higher than expected from such gravitational heating, indicating additional heat input from the AGN \citep{werner2012,babyk2018}. 

Surprisingly, the inner thermodynamic properties of galaxies of all types from massive spirals, through giant ellipticals, central group and cluster galaxies, can be described by similar profiles \citep{babyk2018}. 
The similarity of the profiles across orders of magnitude in halo mass and jet power is even more remarkable given the fact that the hot gas mass of $\approx 10^9~M_\odot$ within $\lesssim10$~kpc is only $\approx 1$\% of the stellar mass within the same radius. It indicates that the AGN feedback, which maintains an approximate balance between heating and cooling, is gentle. \citet{babyk2018} argue that the AGN heating per gas particle is higher in elliptical galaxies than in central cluster galaxies. Such excess heating may explain why most early type galaxies are red and dead with star formation rates below 0.1 $M_\odot$ yr$^{-1}$ while some central cluster galaxies are forming stars. 

\section{Accretion onto Supermassive Black Holes in Elliptical Galaxies}

\subsection{Switching from Radiatively Efficient to Mechanically Efficient Accretion}
\label{sec:switch}

The so-called Soltan argument \citep[][]{1982MNRAS.200..115S} suggests that most supermassive black holes (SMBHs) have accumulated their masses via radiatively efficient accretion. However, most black holes observed at $z\approx 0$ in the centers of giant elliptical galaxies are faint in optical, UV, and X-ray. This is true despite strong evidence that their energy output is able to keep the gas hot even in the most massive clusters. The faintness of SMBHs at the present epoch against the requirement they provide an average energy input of  $10^{44}-10^{45}~{\rm erg~s^{-1}}$ in massive clusters appears contradictory. 

This contradiction can be resolved by two different scenarios. In the first scenario, black holes  are dormant most of the time at $z\approx  0$.  They are active for short periods when their accretion rates are high and most of the energy is released, either mechanically or radiatively. In the second scenario, the mass accretion rate is small, but quasi-continuous.  The SMBH experiences radiatively-inefficient accretion, as predicted by a broad class of models. We discuss  the second hypothesis below, which postulates that the heating rate of the gas is described by the following simple expression \citep[][]{2005MNRAS.363L..91C}
\begin{eqnarray}
H(\dot{M},M_{\rm BH})=c^2 \dot{M} \left [ \alpha_{\rm r} \epsilon_{\rm r}({\dot m})+\alpha_{\rm m} \epsilon_{\rm m}({\dot m})\right],
\label{eq:accretion}
\end{eqnarray}
where $\dot m=\dot M/\dot M_{\rm Edd}(M_{\rm BH})$. The functions $\epsilon_{\rm r}(\dot{m})$ and $\epsilon_{\rm m}(\dot{m})$ characterize the
efficiency of the transformation of the accreted  mass  into radiation and mechanical energy, respectively.  The constants $\alpha_{\rm r}$
and $\alpha_{\rm m}$ are the AGN-gas coupling constants, i.e., they specify the fractions of the
SMBH radiative and mechanical outputs that are eventually dissipated as heat. The key assumption here is that $\alpha_{\rm m}\gg \alpha_{\rm r}$.å In the conditions relevant for present epoch ellipticals, $\alpha_{\rm r} \approx  O(10^{-4})$ \citep[e.g.][]{2004MNRAS.347..144S,2005MNRAS.358..168S}, while $\alpha_{\rm m}\approx  O(1)$ (see \S\ref{sec:keephot}). The direct implication of this assumption is that in order to compensate for gas cooling losses one needs much smaller accretion rates in the regime where most of the energy is released mechanically.

\begin{figure}
\begin{center}
\includegraphics[width=\columnwidth]{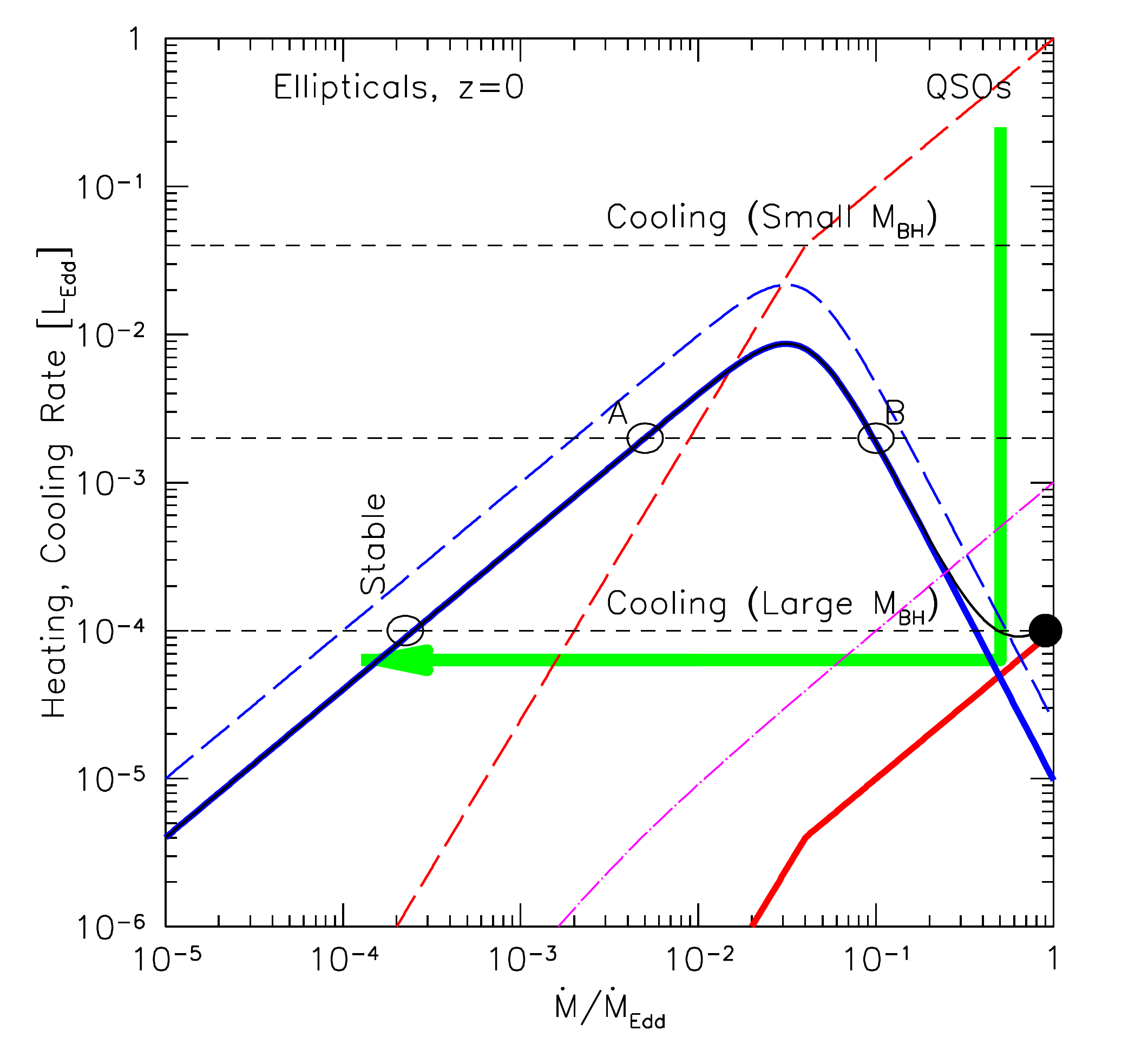}
\end{center}
\caption{Possible evolutionary track of a gas-SMBH system from redshift $z\approx  2-3$ to 0, for the heating rate, described by eq.\,\ref{eq:accretion} \citep[adapted from][]{2005MNRAS.363L..91C}. The horizontal axis gives the mass-accretion rate in Eddington units, while the vertical axis gives the energy release, heating and cooling rates in the same units. The colored dashed curves show the radiative ($\propto \dot{m}\times \epsilon_{\rm r}(\dot{m})$; red) and mechanical ($\propto \dot{m}\times \epsilon_{\rm m}(\dot{m})$; blue) energy releases as a function of $\dot{m}$, respectively. The corresponding radiative and mechanical heating rates (the thick red and blue lines) can be obtained by multiplying these curves by $\alpha_{\rm r}$ and $\alpha_{\rm m}$, respectively. The total heating rate is shown with the black solid line. For comparison, the horizontal dashed lines show the gas cooling rate in the Eddington units (from top to bottom for a small, intermediate and large black hole masses). 
The non-monotonic behavior of the heating rate suggests the following evolutionary scenario. When the mass of the black hole is too small (upper horizontal line),  the feedback from the black hole is not able to compensate for gas cooling losses and the black hole is in the QSO stage with a near-critical accretion rate, high radiative efficiency, and weak feedback. As the black hole grows, it moves down in this plot. The black solid dot marks the termination of this stage, when the black hole is first able to offset gas cooling, despite the low
gas heating efficiency. The lower cooling line illustrates present-day
ellipticals: a stable solution exists at low accretion rates when mechanical feedback from the black hole compensates gas cooling losses. The radiative efficiency of accretion is very low and the black hole growth rate is very slow, yet the system finds itself in point ``A'', where the cooling is balanced by heating, while the AGN appears as a faint source. The major uncertainty in this plot is the mechanical heating rate. For instance, the magenta dash-dotted line shows the mechanical heating rate adopted in \citet[][]{2017ApJ...835...15C}, for which the black hole has to accrete more mass in order to offset gas cooling.}
\label{fig:switch}
\end{figure}

The largest uncertainties in the problem are associated with the functions $\epsilon_{\rm r}(\dot{m})$ and $\epsilon_{\rm m}(\dot{m})$. Theoretical arguments  \citep[e.g.][]{1977ApJ...214..840I,1982Natur.295...17R,1994ApJ...428L..13N} suggest that for accretion rates much smaller than the Eddington rate, i.e., $\dot{m}\ll1$, the radiative efficiency, $\epsilon_r(\dot{m})$ decreases strongly below the canonical value of $O(0.1)$ for radiatively efficient flows, for $\dot{m}\approx  0.01-1$ \citep[e.g.][]{merloni2003}. The dependence of $\dot{m}\times\epsilon_{\rm r}(\dot{m})$ on $\dot{m}$ motivated by these arguments is shown in Fig.~\ref{fig:switch} by the dashed red line. As for the mechanical energy release, a simple ad hoc assumption that the total energy release efficiency $\epsilon_{\rm m}(\dot{m})+\epsilon_{\rm r}(\dot{m})\approx   {\rm const}$ was adopted in \citet[][]{2005MNRAS.363L..91C} (see blue dashed line  in Fig.~\ref{fig:switch} showing $\dot{m}\times \epsilon_{\rm m}(\dot{m})$). Once $\epsilon_{\rm r}(\dot{m})$, $\epsilon_{\rm m}(\dot{m})$, $\alpha_{\rm r}$, $\alpha_{\rm m}$ are specified, their contributions to the total heating rate can be estimated from eq.\,(\ref{eq:accretion}) as shown by the thick red and blue lines in Fig.~\ref{fig:switch}). The total heating rate (the sum of the radiative and mechanical contributions) is shown by the solid black line, which overlaps with the thick blue line at highly sub-Eddington accretion rates.

The non-monotonic behavior of the black curve suggests that for sufficiently massive black holes two solutions exist where AGN heating compensates for the gas cooling losses (see points ``A'' and ``B'' in Fig.~\ref{fig:switch}). The solution ``A'' corresponds to a slow, radiatively inefficient accretion flow and correspondingly slow growth of the black hole.  This is likely what is happening in present day ellipticals, groups and clusters. The solution ``B'' provides the same heating rate, but at much larger accretion rate, with a much brighter AGN, and a rapidly growing black hole \citep[see ][for a discussion]{2005MNRAS.363L..91C}. 
If the mass of the black hole is too small for a given rate of gas cooling losses, there is no solution with $\rm heating \approx   cooling$ at any accretion rate (see the uppermost horizontal dashed line that shows the cooling rate in the black hole Eddington units). We assume that in this case, the gas cools unchecked, the accretion rate is very high, and the black hole grows rapidly. It is tempting to associate this regime with the QSO-like phase of SMBH evolution. This phase continues until the BH accumulates sufficient mass or the gas cooling rate in the system drops substantially.  Then even for a small radiative coupling constant $\alpha_{\rm r}$ the gas heating overcomes cooling losses and the black hole switches to the low accretion rate (and slow growth) regime of type ``A''. There it is nevertheless able to compensate for the gas cooling losses due to the high AGN-gas coupling efficiency $\alpha_{\rm m}$.

\begin{figure}
\begin{center}
\includegraphics[width=0.8\columnwidth]{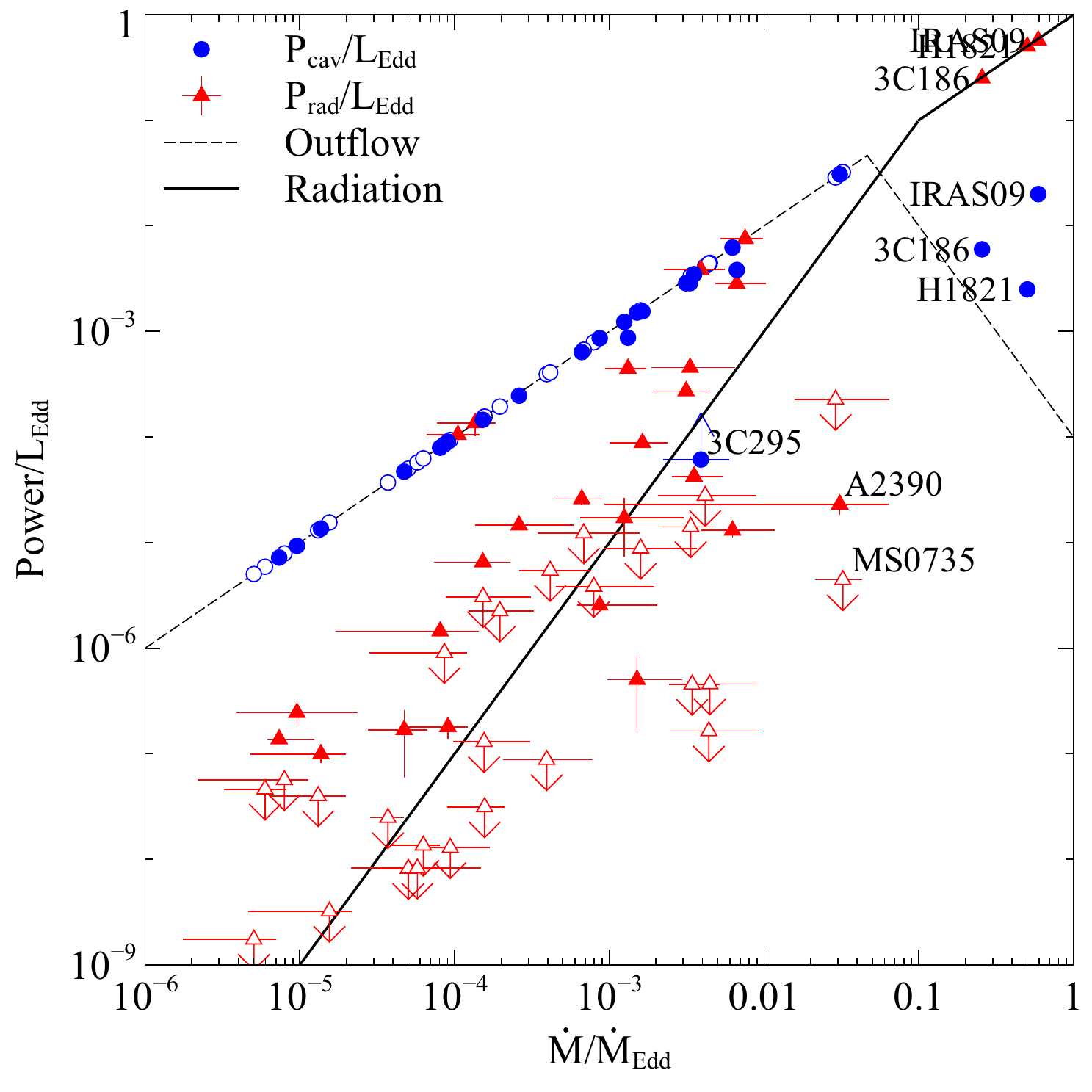}
\end{center}
\caption{The required mean accretion rate scaled by the Eddington rate, $\dot{M}/\dot{M}_{\rm Edd}$, plotted against the cavity power (blue circles) and radiative power (red triangles) scaled by the Eddington luminosity \citep[from][]{russell2013}.}
\label{fig:russell}
\end{figure}

To summarize, in this model, the phase of radiatively efficient accretion and rapid SMBH growth at early times is naturally followed by a radiatively inefficient phase at late times, when a combination of a much lower accretion rate is compensated by efficient coupling of mechanical energy output to the gas. This scenario relies on the assumptions that $\alpha_{\rm m}\approx  O(1)$ and  $\epsilon_{\rm m}(\dot{m})\approx   0.1$ for $\dot{m}\ll 1$. If different prescriptions are used, the system may evolve differently. For instance, \citet{2017ApJ...835...15C} adopted smaller efficiency of the mechanical energy release, shown with the magenta  dash-dotted line in Fig.~\ref{fig:switch}, which lies much below the thick blue line. This means that higher accretion rate is needed in the Ciotti et al.  model to prevent cooling of the same gaseous atmosphere.

This scenario is supported by observation, as summarized in Fig.\ \ref{fig:russell}. Systems that may be in transition between the radiative and mechanical mode have been identified at the centers of several clusters \citep{russell2010,siemiginowska2005,siemiginowska2010,russell2013}.  These systems have extensive X-ray cavities, indicating mechanical feedback occurring over the past hundred million years or so. At the same time they host active quasars. Because the quasar emission is prompt compared to the much older, mechanically-powered bubbles, the data suggest that accreting massive black holes switch modes as the specific accretion rate transitions between the radio and quasar modes. 

Assuming the nuclear black hole masses can be inferred from the host galaxy's luminosity and stellar velocity dispersion, the Eddington ratio may be estimated from the accretion rate derived from the sum of the AGN's radiative and mechanical power. Observation shows that as the specific accretion rate rises, the ratio of  the AGN's radiation power to mechanical power rises \citep{russell2013}. AGN power becomes dominated by radiation when $\dot m/\dot m_{\rm Edd}\geqslant0.01$, as suggested by \citet{2005MNRAS.363L..91C}. However, this behavior is seen only in central galaxies of distant clusters. It has not yet been observed, to our knowledge, in a non-BCG at low redshift.

\subsection{AGN with Low Accretion Rates in Giant Ellipticals}

In principle, the hot atmospheres of giant ellipticals naturally provide a steady supply of fuel to the AGN in their centres. 
Assuming a matter to radiation conversion efficiency of $\epsilon\approx 10$\%, the observed AGN luminosities in ellipticals are usually several orders of magnitude below the predictions of the simple Bondi model \citep{bondi1952}.  However, most of the accretion power in giant ellipticals is likely converted into jets, the power of which inferred from the observed X-ray cavities is comparable to the Bondi values, within a factor of a few \citep[e.g.][]{bohringer2002,churazov2002,dimatteo2003}. 

\citet{allen2006} used {\it Chandra} observations of nine nearby giant elliptical galaxies to determine the Bondi accretion rates calculated from the observed atmospheric gas temperature and density profiles. Black hole masses were inferred from optical velocity dispersion measurements.  Jet energy and power were measured from the pV work and time-scales required to inflate X-ray cavities in the surrounding hot atmospheres. \citet{allen2006} found a tight, power law correlation between the Bondi accretion and jet powers.
Their result indicates that a significant fraction, a few percent, of the rest mass energy of the accreted material emerges in jets. Moreover, the existence of the correlation would suggest that despite the likely presence of magnetic pressure and angular momentum in the accreting gas, the Bondi formulae provide a reasonable description of the accretion process in these systems. 

Such measurements are, however, difficult. The Bondi radius of most systems in the \citet{allen2006} sample are unresolved by {\it Chandra}. Gas density and temperature measurements involve extrapolations across a radial range of nearly a decade.  Furthermore,  the estimates of jet power based on radio data and X-ray imaging also have factors of several uncertainties. Using a sample of 12 nearby systems, \citet{russell2013} did not confirm the correlation.

The X-ray emitting hot gas within the Bondi radius can currently only be resolved in a handful of systems, which include NGC~3115 \citep{wong2011,wong2014} and M~87 \citep{russell2015,russell2018}.
Very deep (1 Ms) {\it Chandra} observations of NGC 3115 indicate shallow density profiles of $\rho\propto r^{-1}$ and the presence of multi-phase gas spanning a temperature range of 0.3--1 keV at the Bondi radius of the central black hole. The coolest temperature component is located in the central 150 pc; it may be circulating and cooling toward a disk region. For a constant inflow rate the expected density profile is  $\rho\propto r^{-1.5}$. Outflows, would decrease the inflow rate with the decreasing radius \citep[e.g.][]{yuan_narayan2014}, which would flatten the density distribution to $\rho\propto r^{-0.5-1}$. The observed shallow density profile is consistent with most of the inflowing material being ejected before it reaches the event horizon.

Perhaps the most illuminating observational study of the accretion of the hot X-ray emitting galactic atmosphere onto the central AGN has been performed on M~87 by \citet{russell2018}. The mass of the supermassive black hole in M~87, estimated using the analyses of gas-dynamics and stellar-dynamics, is $M_{\rm BH} = 3-6\times10^9 M_\odot$ \citep{gebhardt2011,walsh2013}.  This corresponds to a Bondi radius $r_{\rm B} = 0.12-0.22$~kpc (1.5--2.8 arcsec). 
The dedicated 300 ks {\it Chandra} observation, carried out with a short frame time of 0.4 s and using a subarray to minimise pileup from the bright jet, resolves the density profile inside $r_{\rm B}$ in three azimuths.  
The density gradient is steep in sectors to the N and S (perpendicular to the jet), with $\rho\propto r^{-1.5\pm0.1}$, and significantly shallower along the jet axis to the E, where $\rho\propto r^{-0.93\pm0.07}$. This density structure is consistent with steady inflows perpendicular to the jets and an outflow along the jet axis. However, \citet{russell2018} argue, that the actual inflow speed $v_{\rm r} < 8\pm2$ km~s$^{-1}$, which rules out the Bondi accretion. The gas flow is subsonic and must be supported by pressure or rotation. The estimated spherical mass inflow rate $\dot{m}_{\rm acc}<0.010\pm0.003~M_\odot$ yr$^{-1}$, which is at least an order of magnitude less than the the Bondi rate of 0.1--0.5 $M_\odot$ yr$^{-1}$. Assuming an efficiency of 10\%, this accretion rate limit could still supply about an order of magnitude more power than is required for the observed AGN activity. However, the majority of the inflowing material is either blown away by the outflow along the jet or may be consumed by star formation \citep[see][]{mcnamara1989} before it reaches the black hole. 

\citet{russell2018} also point out that within the gravitational sphere of influence of the central supermassive black hole in M~87, the hot gas is multiphase and spans temperatures from 0.2 to 1 keV. The radiative cooling time of the lowest temperature gas drops to only 0.1--0.5 Myr, which is comparable to its free fall time. The multi-phase hot atmosphere at the centre of M~87 most probably cools catastrophically within the Bondi radius to form a mini cooling flow. 

Rapid cooling should produce in dense, cool, gas condensations that decouple from the hot atmosphere. If this cooling gas has an angular momentum then it will feed into the cold gas disk within the innermost $r\approx 80$ pc around the nucleus \citep{ford1994}. Strong limits on the current accretion rate in a geometrically thin disk are found by \citet{prieto2016}. Some of the coolest X-ray emitting gas is spatially coincident with optical emission line nebulae, which might trace even colder gas phases. Narrow band H$\alpha$+[NII] images  suggest that some ionized gas might also be reaching the nucleus from larger radii in inflows approximately perpendicular to the jets \citep{ford1979,sparks1993,werner2010,boselli2018}. While the steady hot inflow of volume-filling gas might be sufficient to power the jets seen today in M~87, clumpy cooler gas with a small volume filling fraction could reach the core episodically, potentially triggering significantly larger outbursts. 

These approximate conditions may pertain to other giant elliptical galaxies. While accretion from hot atmospheres would provide ample fuel to most low power, relatively quiescent giant ellipticals, even at substantially reduced accretion rates with respect to the Bondi rate, it is insufficient to power the strong radio-jets in clusters exceeding $\approx 10^{45}$~erg~s$^{-1}$ \citep{mcnamara2011,nemmen2015}. These systems may require a supplementary inflow from cold gas accretion \citep{pizzolato2005,gaspari2013} or additional power from black hole spin energy \citep{mcnamara2011}.

\section{How do AGN Heat Galactic Atmospheres?}
\label{sec:keephot}

\subsection{Heating by Radiatively Efficient AGN}

In this section we outline a basic physical picture of the heating of galactic and cluster atmospheres by AGN \citep[see, e.g.\ ][for recent reviews]{2012ARA&A..50..455F,2012NJPh...14e5023M,2014PhyU...57..317V,2016NewAR..75....1S}. We do not consider the role of supernovae and stellar winds, which become progressively more important as the mass of the halo decreases \citep[see, e.g.][for a recent review]{2017ARAA..55...59N} and we do not discuss the very rich field of numerical modeling of the AGN-gas interaction. 

Elliptical galaxies represent an important class of objects in-between galaxy clusters and normal galaxies. As such they offer us a possibility to extend the lessons on the AGN feedback from galaxy clusters to the less massive systems and potentially to all spheroids. At the same time, there are significant differences between elliptical galaxies and clusters that allow one to test the AGN feedback models in more extreme conditions. These differences include i) larger ratio of the black hole mass to the total atmospheric gas mass and ii) lower virial temperature of the atmosphere. The former difference implies that the energy released by the black hole could potentially expel the gas from the entire galaxy, while the latter allows for a more rapid cooling of the metal-rich gas, especially in low-mass ellipticals. It appears that both factors play a role, depending on the mass of the system and its evolutionary history. 

Accretion of matter onto a SMBH can produce copious amounts of radiation and powerful jets/outflows, which in terms of energy release greatly exceed the gas radiative cooling losses in individual galaxies. Therefore, both flavors of the AGN feedback (radiative and/or mechanical) could be considered.

The radiation couples to the gas via photoionization of electrons and by interactions of photons with electrons via the Compton effect. The Compton temperature of electrons $T_{\rm C}$ depends only on the shape of the AGN spectrum and can be of the order $2\times10^7~{\rm K}$ \citep[][]{2004MNRAS.347..144S}. This value is higher than the virial temperature of a giant elliptical galaxy, implying that the radiative output of the AGN  can drive the gas out of the galaxy if the gas cooling losses are small enough, so that the gas temperature can attain $T_{\rm C}$. 
For the present-day ellipticals, the overall efficiency of capturing the radiation is not high, partly due to the small Thomson depth of the gas in ellipticals and due to inefficient energy transfer from photons to electrons. However, during powerful (radiative) AGN outbursts the impact of the radiation can still be significant  \citep[e.g.][]{2005MNRAS.358..168S,2007ApJ...665.1038C,2017ApJ...835...15C}.  The theory of accretion onto black holes suggests that in order to have high radiative efficiency the mass accretion rate should not be very far from the Eddington level $\displaystyle \approx  2 M_\odot \;{\rm yr^{-1}} \left ( \frac{M_{BH}}{10^8\; M_\odot}\right )$. In the radiative feedback models, the AGN spends only a fraction of time in a radiatively-bright high-accretion-rate mode, while most of the time the accretion rate is  small and the AGN is faint. The overall AGN-gas coupling efficiency in the present-day giant ellipticals is relatively low and most of the released energy escapes the system.

\subsection{Mechanical Heating by Radio AGN}

The mechanical energy released by the AGN can have many orders of magnitude higher coupling efficiency with the gas and therefore allows for a quasi-steady solution at much lower accretion rates \citep[e.g.][]{2005MNRAS.363L..91C}. The energetics of jets and outflows from radio-bright AGNs living in elliptical galaxies (either in BCGs or in isolated galaxies) has long been appreciated \citep[e.g.][]{1990MNRAS.246..477P,1996ApJ...467..597B,1996MNRAS.283..873R}.  

Early suggestions of the impact of SMBHs on the gas in ellipticals include, e.g. \citet{1995MNRAS.276..663B,silk1998}. In \citet{1995MNRAS.276..663B} the role of jets was to prevent catastrophic cooling in the core of the galaxy by episodic outbursts of AGN activity, but not necessarily leading to reheating of the entire atmosphere, 
while \citet{silk1998} argued for a more powerful outflow that is able to accelerate the entire baryonic gaseous atmosphere to the escape velocity of the halo. Effectively, the latter scenario requires an amount of energy comparable to the thermal energy of the gas to be released during the sound crossing time of the system. 

Observations of nearby galaxy clusters suggested yet another scenario, in which AGN provide a quasi-steady injection of energy that on average compensates the gas cooling losses and maintains a long-lived atmosphere in quasi-equilibrium. A combination of radio and X-ray data revealed a close correspondence between radio-bright lobes produced by jets and depressions in the X-ray surface brightness \citep[e.g.,][]{1993MNRAS.264L..25B,1995MNRAS.274L..67B,1995ApJ...447..559S}, showing that AGN do perturb the gas. When the energy is suddenly injected by jets into an unperturbed gaseous atmosphere (assuming for simplicity that the energy injection is isotropic) it drives a strong shock that compresses and heats the gas \citep[e.g.][]{1998ApJ...501..126H}. However, when the energy input is steady then the expansion eventually becomes subsonic and most of the energy does not go into heat, but is stored as the enthalpy of the inflated bubble $H=\displaystyle \frac{\gamma_b}{\gamma_p-1}P_bV_b$, where $V_b$ is the volume of the bubble, $P_b\approx   P_{\rm gas}$ is the pressure inside the bubble and $\gamma_b$ is the adiabatic index of the medium inside the bubble ($\gamma_b=4/3$ if the bubble is filled with relativistic plasma). The configuration with a hot and low density bubble located at the bottom of the potential well filled with colder/denser gas is of course unstable \citep[e.g.][]{1973Natur.244...80G}. Eventually the motions induced by buoyancy forces overcome the expansion velocity and the bubble starts rising in the potential well. 

\begin{figure*}
\includegraphics[width=0.464\columnwidth]{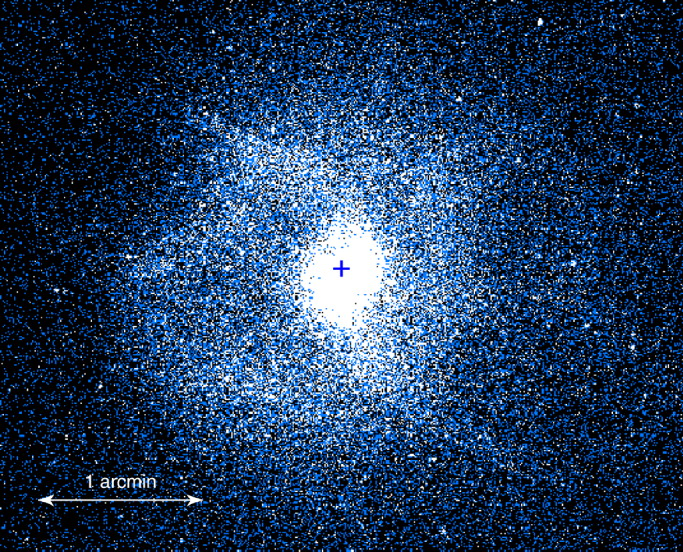}
\includegraphics[width=0.49\columnwidth]{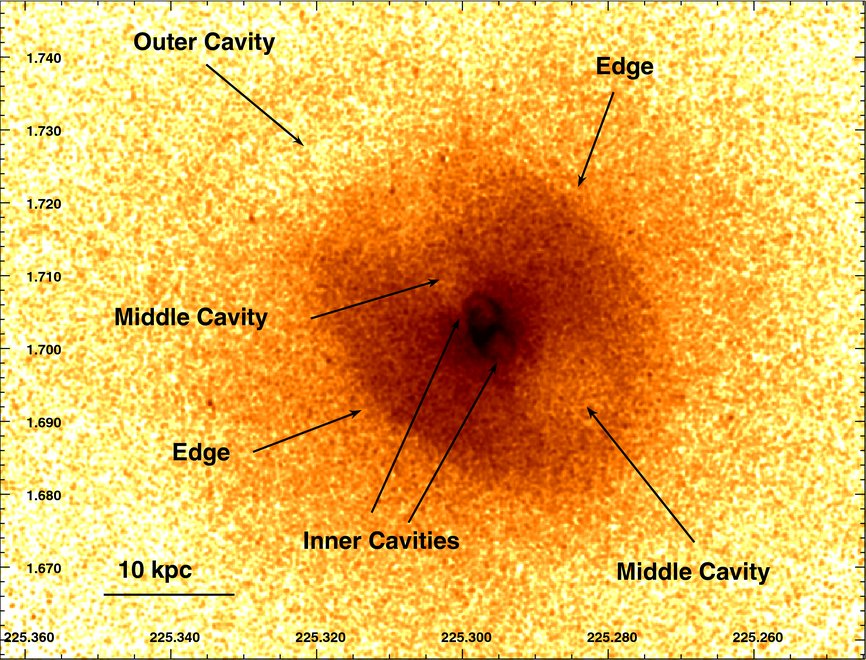}
\caption{X-ray images of NGC~4636 \citep[left, from][] {2002ApJ...567L.115J} and NGC~5813 \citep[right, from][] {randall2015}, showing X-ray emitting gas being strongly perturbed by the AGN energy release.}
\label{fig:shocks}
\end{figure*}

These arguments  provided the first robust estimates of AGN mechanical power in the Perseus and Hydra~A clusters \citep[][]{churazov2000,mcnamara2000}. The derived power turned out to be comparable to the total gas cooling losses. This conclusion has been confirmed and extended for objects spanning a wide range of masses and cooling luminosities, including elliptical galaxies \citep[e.g.][]{2004ApJ...607..800B,2012MNRAS.421.1360H}. Despite substantial (and inevitable) uncertainties in the estimated power this exercise shows that jets/outflows can provide enough power to keep the gas hot.
We note here that  the above procedure yields a lower limit on the released energy. Indeed, when radiative losses and leakage are negligible, the energy required to inflate a radio lobe is the sum of its internal energy and the $pdV$ work it does on its surroundings.  Using the enthalpy estimates the latter as $pV$.  The total $pdV$ work generally exceeds $pV$ because the pressure the lobe expands against decreases with time. In part, that may be because the early expansion is supersonic (so that the lobe pressure exceeds the pressure of the unperturbed gas) and the pressure in a stratified atmosphere is a decreasing function of the radius. Similar clear signs of a powerful mechanical feedback are seen not only in clusters, but in groups and massive early type galaxies too. Two spectacular examples of elliptical galaxies heavily affected by activity of the SMBHs are shown in Fig.~\ref{fig:shocks}.

\subsection{AGN Energy Dissipation in the Atmosphere}

Let us reiterate that in the quasi-continuous injection scenario, most of the energy goes into the enthalpy of the inflated bubble, rather than into immediate heating of the gas, as sometimes stated. The energy is stored instead as the thermal energy of the bubble $\displaystyle E_t=\frac{1}{\gamma_p-1}P_bV_b$ and the potential energy of the displaced gas $\displaystyle E_p=P_bV_b$. However, as the bubble rises through the atmosphere, all this energy will be released to the ambient thermal gas \citep[][]{churazov2001,churazov2002,2001ASPC..240..363B}. Different physical processes could  contribute to the energy transfer, including generation of turbulence in the wake of the bubble; viscous dissipation; uplift of the low entropy gas, which is entrained by the rising bubble; excitation of internal waves, etc. We return to this question at the end of this Section. However, the basic conclusion that the coupling efficiency of the buoyantly rising bubbles to the gaseous atmosphere is close to $100\%$ holds, since it is governed by energy conservation. 

This is clear from the following simple consideration. Imagine an essentially massless (filled with relativistic particles) bubble rising in an atmosphere at a steady terminal velocity. The buoyancy force has to be compensated by a drag acting on the bubble $F_{\rm drag}\approx  F_{\rm buoyancy}=g\rho_{\rm gas}V_{\rm b}$. It means that the work done by the bubble while rising along the radius by  $\Delta r$, $E\approx  g\rho_{\rm gas}V_b\Delta r$ is equal to the change of the bubble potential energy. That would be the case for a bubble whose volume does not change as it rises. In fact, if the bubble stays in pressure equilibrium with the ambient gas, then the volume $V_{\rm b}$ changes accordingly, extracting thermal energy from the relativistic particles so that this energy is also used. In other words, the energy gained by the gas near the bubble equals the decrease in the bubble enthalpy (this is simply energy conservation for the internal energy, while the additional $pV$ in the enthalpy represents gravitational potential in the atmosphere that is also released to the gas). Therefore, for an adiabatic bubble with pressure $p$, the enthalpy is proportional to $p^{(\gamma_b - 1)/\gamma_b}$.  For $\gamma_b = 4/3$ (bubble filled with relativistic particles), the exponent of the pressure is 0.25, so that pressure would need to decrease by a factor of 16 for the bubble to transfer half the available energy to the gas \citep[see Fig, 13 in ][]{forman2017}.  This is equivalent to crossing $\log 16 \approx 2.8$ scale heights (and less than 2 scale heights for $\gamma_b = 5/3$).

Even after the energy is transferred to the gas, it might need to go through a few more steps before it is dissipated as heat (e.g., 
bubbles $\rightarrow$ turbulent gas motions $\rightarrow$ dissipation). Despite these extra steps, there are good reasons to believe that in the majority of plausible scenarios only a small fraction of energy escapes the system. For instance, internal waves that could be excited in stratified atmospheres are trapped because of the decline of the buoyancy frequency with radius \citep[][]{1990ApJ...357..353B}. The same is true (for different reasons) for many other forms of perturbations generated by the AGN.  Therefore one can make an almost assumption-independent statement, that all energy provided by jets and outflows will eventually be dissipated in the gas and will not escape from the system, thus ensuring high efficiency of the mechanical AGN feedback. 

The above statement does not eliminate the need to identify physical processes that lead to the dissipation of the energy provided by AGN. To answer this question from first principles one needs comprehensive models of weakly collisional plasma, including kinetic effects, and of the jet, which are not yet available. Therefore various simplifying assumptions are adopted, making this field an active area of research.  

\subsection{Heating by Strong Shocks}

One obvious way of heating the gas is by driving a strong shock through it. Shocks, driven by AGN activity, are indeed observed in many elliptical galaxies \citep[e.g.][]{2002ApJ...567L.115J,baldi2009,machacek2006,2007ApJ...665.1057F,randall2015}. Clear examples of multiple shocks (albeit weak ones) in NGC~4636 and NGC~5813 are shown in Fig.~\ref{fig:shocks}. The shock heating is a direct way of increasing the gas entropy, provided that the Mach number is high enough, since for weak shocks the increase of entropy at the shock front scales as the pressure difference between the pre- and post-shock regions to the 3rd power. In contrast to the quasi-steady energy-injection scenario, outlined above, the strong-shock model would reach the highest efficiency if the energy is released in short outbursts, separated by quiescent periods, long enough so that the shock-heated gas is evacuated by buoyancy from the central region and is replaced by the fresh low-entropy gas. Strong shocks can be associated, for example, with the momentum-dominated outflows that produce shock-heated gas at the jet/outflow termination point and form a cavity filled with this gas \citep[e.g.][]{2003ApJ...596L..27K}.  Efficient shock heating can also be achieved in a scenario with a spherically symmetric energy-driven outburst, provided that a large amount of energy $E$ is released from a small region in a time interval much shorter than the sound crossing time of a volume containing an amount of thermal energy comparable to $E$ \citep[see, e.g.][]{1998ApJ...501..126H,2015ApJ...815...41R,2017MNRAS.468.3516T}. This corresponds to an ``instantaneous'' explosion when about 90\% of the released energy goes into shock-heating \citep[][]{2017MNRAS.468.3516T}. 

Alternatively, one can assume that the AGN is directly supplying very hot, nonrelativistic plasma. Let us assume that i) the net outcome of the AGN activity is the formation of cocoons/cavities (or strongly shocked envelopes), in which energy density is dominated by very hot (but nonrelativistic) plasma and ii) this plasma can mix with the ambient (lower entropy) gas. This case corresponds to the ``entrainment and mixing'' scenario \citep[e.g.][]{2002Natur.418..301B,2004MNRAS.348.1105O,2005ApJ...622..847S,2006ApJ...643..120B,2012MNRAS.424..190G,2015ApJ...815...41R,2016MNRAS.455.2139H}. Once  the hot and ambient gases are mixed on small scales, Coulomb collisions redistribute the energy between all particles. The ability of very hot particles to share their energy with the rest of the gas is an important ingredient of the mixing scenario, implying that these particles (at least the protons) are not relativistic. In principle, the presence (or the lack) of thermal gas inside the cavities could be tested with SZ-effect observations \citep[e.g.][]{2005A&A...430..799P,2010A&A...520A.106P,abdulla2018}. The question whether the cavities mix with the ambient gas on microscopic scales, or the interface is somehow stabilized is still open \citep[e.g.][]{2005MNRAS.357..242R,2007MNRAS.378..662R}. If mixing happens only on macroscopic scales then the heating is not due to the energy exchange between particles and other mechanisms (discussed below) are responsible for the eventual dissipation. 

\subsection{Heating by Weak Shocks \& Sound Waves}

A more gentle version of heating could be achieved with weak shocks \citep[e.g.][]{2006ApJ...638..659M,randall2011}, which does not require mixing, because the heating is distributed and particles change their energies by a small factor and the redistribution of energy is not required. However, some fine-tuning is needed if one wants to channel most of the energy, released by the AGN into weak shocks. Even more gentle is the scenario of heating the gas with sound waves \citep[][]{2003MNRAS.344L..43F,2006MNRAS.366..417F,fabian2017}. The heating by sounds waves relies on conduction or viscosity as the energy dissipation mechanism, operating when the waves  propagate radially from the central region \citep[][]{2004ApJ...611..158R,2006MNRAS.366..417F,2018ApJ...858....5Z}. Once again, an important question is how to channel a significant fraction of the AGN energy into sound waves. For an instantaneous spherically-symmetric outburst this fraction does not exceed $\approx $12.5\%  and is smaller for a quasi-continuous explosion \citep{2017MNRAS.468.3516T}, although it is plausible that for a collimated momentum-dominated jet/outflow this fraction can be larger.

\subsection{Cosmic Ray Heating}

Yet another ``mixing'' scenario involves cosmic rays. If most of the AGN energy is first stored in the form of relativistic protons, which can mix with the ambient gas (or leak) from lobes/cocoons \citep[e.g.][]{2003A&A...399..409E,2008MNRAS.383.1359R}, then the question arises how those protons couple with the thermal gas, given that the lifetime of relativistic protons (CRps) is longer than the Hubble time \cite[e.g.][]{1996SSRv...75..279V}. There are three possibilities: i) protons remain ``frozen'' in the gas, ii) they (quickly) diffuse through the gas outside the dense part of the atmosphere, or iii) they stream (move with respect to the gas frame) and collectively excite plasma waves that are continuously dumped into heat.  The first two scenarios do not imply any heating at all. The main constraints on these two scenarios come from the lack of gamma-ray emission due to $\pi^0$ decay \citep[e.g.][]{2014ApJ...787...18A,2014A&A...567A..93P}, suggesting that either protons diffuse quickly or they remain confined within the bubbles \citep[][]{2017MNRAS.470.3388P}. It is still possible however that the mixing does not go all the way to microscopic scales, but the protons remain confined to tiny bubbles, which themselves are ``frozen'' to the gas. In this case, the gamma-ray limits are not relevant, since relativistic protons are isolated from thermal protons, and the role of CRps is to make the gas lighter and lift it up to an altitude in the atmosphere where the effective entropy of the gas is similar \cite[][]{churazov2000}. Finally, the 3rd scenario, where the CRps stream through the gas, can lead to a net gas heating \citep[e.g.][]{1988ApJ...330..609B,1991ApJ...377..392L,2008MNRAS.384..251G,2011ApJ...738..182F,2013MNRAS.434.2209W,2013ApJ...779...10P,2017ApJ...844...13R}. The energy losses due to streaming can alleviate the gamma-ray constraints if the protons lose their energy due to streaming fast enough. This scenario 
is also not free from assumptions related to the rate of CRps mixing (or leakage from the bubbles) and plasma physics aspects of streaming, in particular, the dumping mechanisms that let protons move fast enough.

\subsection{Coupling Bubble Enthalpy to the Hot Atmosphere}

We now return to the case of a bubble, created by a quasi-continuous AGN jet/outflow, which is fully isolated from the ambient gas so that neither particles nor energy can flow through the boundary. This is an ad hoc assumption on par with the converse assumption that the bubbles can easily mix with the ambient gas. 
In this case it does not matter whether the bubble is filled with relativistic or non-relativistic plasma\footnote{It matters only when one needs to estimate the thermal energy inside the bubble, which for a given volume and pressure depends on the adiabatic index of the gas.}, as long as the particles inside the bubble are much hotter than those outside. As discussed above, the energy exchange between such bubbles and the surrounding medium is purely mechanical and is happening during the buoyant rise of the bubbles. Among the mechanisms that provide the coupling of the bubbles to the gas are: excitation of waves, viscous stresses, excitation of turbulence in the wakes, entrainment and uplift of low entropy gas, etc. \citep[e.g.][]{churazov2001}. Sound waves cannot be efficiently excited if the rise velocity of the bubble is steady and subsonic, while internal and Alfven waves can, although to become a dominant process of the energy extraction from the bubble, or, equivalently, the dominant contributor to the drag acting on the bubble, these mechanisms have to win against other processes. For instance, \citet[][]{2018MNRAS.tmp.1212Z} have shown that if the bubble is flattened in the vertical (along the radius) direction, then the contribution of internal waves can be significant, or even dominant. That would fit well into the turbulence-heating scenario \citep[e.g.][]{2005ApJ...622..205D,2014Natur.515...85Z}, where the turbulence is excited either directly by the rising bubbles or by the non-linear evolution of internal waves, which themselves are excited by rising bubbles. However, if the bubbles are able to mix even in the presence of magnetic fields, the excitation of internal waves can become inefficient \citep[e.g.][]{2015ApJ...815...41R,2016ApJ...829...90Y,2018ApJ...857...84B} compared to other processes. At the same time, the presence of a magnetic field in the gas may itself induce a drag on the rising bubble \citep[e.g.][]{2006MNRAS.373...73L}, which was found in numerical simulations to be larger than the pure hydrodynamic drag \cite[][]{2008ApJ...677..993D}. Overall, it is clear that there are many different (linear or nonlinear) ways of extracting the energy from the bubble even if it is completely isolated from the ambient gas. 

\subsection{Summary}

The above discussion can be summarized as follows:
\begin{enumerate}
\item We have solid observational evidence that AGN activity strongly perturbs hot gaseous atmospheres in early type galaxies.
\item The amplitude of perturbations strongly suggests that the AGN release enough energy to keep the gas hot.
\item We have solid theoretical arguments suggesting that the released energy does not leave the system and eventually dissipates, i.e., the coupling efficiency of the AGN and the gas is very high.
\item So far there is no consensus about the dominant physical mechanisms governing the energy flow through the system and its eventual dissipation.
\end{enumerate}
Thus, there is a very good reasons to believe that AGN keep the gas hot (this is an almost model-independent statement), but there are enough uncertainties in the properties of the jets and the gas to allow for different heating scenarios, which all lead to the same final result. The observational data are not yet sufficient to unambiguously single out the most favorable scenario,  but the situation will improve with high angular resolution SZ measurements and future velocity measurements with {\it XRISM}, {\it ATHENA}, and {\it LYNX}.

\section{Atmospheric Dynamics}
\label{sec6}

As discussed above, X-ray and radio images of giant elliptical galaxies show interactions between the radio plasma injected by AGN and the hot galactic atmospheres. These interactions are expected to induce atmospheric gas motions \citep[e.g.][]{churazov2001}. Atmospheric motions are difficult to detect and measure.  The spectral resolution of current CCD-based detectors are too low.  The spectral lines resolved by the reflection grating spectrometers (RGS) on {\it XMM-Newton} are broadened by the spatial extent of the source. 

All useful X-ray emission lines formed in galactic atmospheres are located in the soft X-ray spectral band. Unfortunately future X-ray calorimeter spectrometers with $\Delta E\approx  5$~eV provide the poorest spectral resolution $\Delta E/E$ in the soft band. Therefore, precise measurements of gas velocities $v\lesssim150$~km~s$^{-1}$ will remain challenging for the foreseeable future. Measurements of gas motions in cool, low-mass systems therefore largely have to rely on indirect methods. 

The first upper limits and measurements of the velocity broadening of X-ray emission lines for giant ellipticals, groups and clusters of galaxies using RGS were performed by \citet{sanders2011b}. This work was later extended by \citet{sanders2013} and \citet{pinto2015}. For more than half of the elliptical galaxies, the measurements of line widths provided a 68\% upper limit $\lesssim200$~km~s$^{-1}$.

Measurements of the suppression of lines with high oscillator strengths by resonance scattering provide an independent way to determine the characteristic velocities of small-scale turbulence \citep[][]{gilfanov1987,churazov2010b}. At the temperatures of giant elliptical galaxies, $kT\lesssim0.9$~keV, the most sensitive spectral lines to determine the level of resonant scattering are the strong \ion{Fe}{xvii} lines at 15.01 \AA\ (2p-3d) and the unresolved blend of the same ion at 17.05 and 17.10 \AA\ (2p-3s). The oscillator strength of the 15.01 \AA\ line is $f=2.49$, and thus is expected to have a relatively large optical depth. However, the oscillator strengths of the 17.05 and 17.10 \AA\ lines are $f=0.126$ and $f=5.2\times10^{-8}$, respectively. Therefore, the optical depth of this line blend is negligible. Because of this dramatic difference in optical depths, and the fact that the intensity ratios of these lines are weakly dependent on temperature, the comparison of their intensities allows us to measure the magnitude of resonant scattering. 

\citet{werner2009} and \citet{deplaa2012} measured the resonant scattering of the 15.01 \AA\ \ion{Fe}{xvii} line in a sample of X-ray bright nearby galaxies.  They  compared their results to models that accounted for resonance scattering under differing values of the characteristic turbulent velocity. \citet{werner2009} found relatively low velocities of $v_{\rm turb}\lesssim100$~km~s$^{-1}$ in NGC~4636. While \citet{deplaa2012} found a turbulent pressure $\approx 30$\% of the thermal pressure in NGC~5813, the inferred velocities in NGC~5044 are surprisingly high, implying a turbulent pressure support of $\gtrsim40$\%. The systematic uncertainties on these results are, however, large due to uncertainties in atomic data. 

 \citet{ogorzalek2017} have done the most detailed systematic study of atmospheric velocities in giant ellipticals using RGS spectra. By combining measurements of resonance scattering and line broadening \citep[based on][]{pinto2015}, they obtained constraints for 13 galaxies. Assuming isotropic turbulence, they obtained a best-fit, mean 1D turbulent velocity of $\approx 110$~km~s$^{-1}$.  This velocity implies a 3D Mach number of $\approx 0.45$ and a non-thermal pressure support of $\approx 6$ per cent.  To within the uncertainties, the non-thermal pressure is negligible. This supports the picture of a quasi-continuous, gentle feedback discussed in previous sections. 

Comparison between gravitational potential profiles for giant ellipticals derived from X-ray and optical data suggests that the combined contribution of cosmic rays, magnetic fields and micro-turbulence to atmospheric pressure is $\approx 20-30$\% of the gas thermal pressure \citep{churazov2010}. Taken at face value, in combination with the
results of \citet{ogorzalek2017}, these measurement would hint at a 10--20\% pressure contribution by cosmic rays and magnetic fields. However, the systematic uncertainties associated with these
measurements are large.

Sub-mm, infrared, and optical spectra of the cold and warm gas phases in giant ellipticals also provides indirect estimates on the velocities in hot galactic atmospheres. Using simulations of AGN feedback, \citet{gaspari2018} predict that all gas phases in giant ellipticals are tightly linked in terms of the ensemble velocity dispersion. The best-fit linear relation between the 1D velocity dispersion of warm cloudlets and filaments condensing out of the hot atmosphere and the 1D velocity dispersion of the hot gas determined from their simulations is  $\sigma_{\rm v, warm}=0.97^{+0.01}_{-0.02}\sigma_{\rm v, hot}+8.3^{+3.5}_{-5.1}$~km~s$^{-1}$.  This indicates that spectroscopic measurements of velocity dispersion in the warm/cold gas within apertures of several kpc are good tracers of the velocity dispersion of the hot X-ray emitting gas. \citet{gaspari2018} also gathered $\sigma_{\rm v}$ and $v_{\rm LOS}$ measurements for the warm and cold gas in 72 giant ellipticals, and brightest group/cluster galaxies \citep[based on measurements reported in][]{mcdonald2012,werner2013,werner2014,hamer2016,temi2018}. Their velocity dispersions are in the range $\sigma_{\rm v}\approx 90-250$~km~s$^{-1}$, with a mean $\approx  150$~km~s$^{-1}$. This range is broadly consistent with constraints from the high-resolution X-ray grating spectra \citep{ogorzalek2017}.

\section{Summary and Outstanding Unsolved Problems}

\begin{itemize}
\item Observation has established that galaxies with halo mass similar to the Milky Way ($\approx 10^{12}~M_\odot$) or higher host hot X-ray emitting atmospheres and central supermassive black holes. \\

\item Simulations also predict that halos with masses of $\approx 10^{12}~M_\odot$ or higher will harbor hot atmospheres. However, the value of the mass above which galaxies hold on to hot atmospheres remains uncertain. Its value may depend on a combination of factors including, redshift, feedback, environment, and dynamics. \\

\item The origin of the hot galactic atmospheres is still in debate. Observation indicates that most of the X-ray emitting gas was accreted externally and heated by shock waves during the process of galaxy assembly. Stellar ejecta that has thermalised and mixed with the accreted material may contribute significantly to the atmospheric gas mass. However, data are inconsistent with a simple stellar mass loss alone. The relative fraction of accreted and internally produced material may vary with total mass and growth history, with accretion dominating at high masses and stellar ejecta in lower-mass, disky galaxies. \\

\item While the total baryon fraction of Milky Way mass ellipticals is roughly consistent with the cosmic value, isolated, massive spirals are apparently missing more than half of the baryons within their virial radii. The reason for this difference is not understood. \\

\item Recent, high-fidelity measurements and models show that the metallicities and relative abundances of $\alpha$ elements with respect to iron in hot atmospheres of ellipticals and groups are remarkably similar to those in massive clusters. The chemical composition of giant elliptical atmospheres is at variance with their stellar populations, which typically show an overabundance of $\alpha$ elements. \\

\item Recent studies have shown that, while some molecular gas in ellipticals may have accreted externally, most has likely cooled from their hot atmospheres. \\

\item Despite the enormous range in halo and atmospheric gas mass between elliptical galaxies and central galaxies in clusters, their entropy and cooling time profiles, which characterize their thermodynamic histories, are remarkably similar. \\

\item While much has been learned in the past decade about the relationship between AGN feedback and atmospheric cooling, much remains to be done. Recent numerical simulations and models of thermally unstable cooling have generated great interest. But a clear picture of how thermally-unstable cooling ensues that can be sharply constrained or ruled-out by observation continues to elude us.  Heating and cooling appears closely balanced in clusters.  However, galaxies may be less stable, as in evidence by the large scatter in their X-ray luminosities.  We do not know how much scatter is driven by intermittent AGN outbursts. Our picture might also be skewed by observational selection: galaxies observed deeply by {\it Chandra} and {\it XMM-Newton} are biased towards the brightest systems in X-rays. Importantly, the role of radio-mode AGN feedback in disky galaxies has yet to be fully probed by observation. \\

\item Observations discussed in this review show that in massive ellipticals, jets emanating from blacks holes accreting at highly sub-Eddington rate are sufficiently powerful to balance the radiative cooling of hot atmospheres and limit further star-formation. The heating rate is tuned to atmospheric cooling. No consensus has been achieved on the dominant mechanism responsible for energy transport from jets, to X-ray bubbles, and eventually into the hot plasma at large. \\
\end{itemize}

\begin{acknowledgements}
NW was supported by the Lend{\"u}let LP2016-11 grant awarded by the Hungarian Academy of Sciences.
BRM thanks the Natural Sciences and Engineering Research Council of Canada and the Canadian Space Agency for financial support.
\end{acknowledgements}
\bibliographystyle{aps-nameyear}
\bibliography{clusters}

\end{document}